\documentclass[journal]{IEEEtran}

\usepackage{threeparttable}
\usepackage{amsmath}
\usepackage[noend]{algpseudocode}
\usepackage{array}
\usepackage{multirow}
\usepackage[linesnumbered,ruled]{algorithm2e}
\usepackage{gensymb}
\usepackage{color}
\usepackage{soul}
\usepackage{fancyhdr}
\usepackage[pass]{geometry}
\usepackage{balance}
\usepackage{textcomp}
\ifCLASSINFOpdf
  \usepackage[pdftex]{graphicx}
\else
\fi
\usepackage{amsmath}

\DeclareMathOperator*{\argmin}{arg\,min}

\usepackage{footnote}
\usepackage{subfig}
\makesavenoteenv{tabular}
\usepackage{tablefootnote}
\usepackage[hang]{footmisc}
\usepackage{subfig}
\usepackage{mathtools}
\DeclarePairedDelimiter{\ceil}{\lceil}{\rceil}
\usepackage{amsmath,amssymb,amsfonts}
\usepackage{cite}
\usepackage{multirow}
\newcommand{\nosemic}{\renewcommand{\@endalgocfline}{\relax}}
\newcommand{\dosemic}{\renewcommand{\@endalgocfline}{\algocf@endline}}
\let\oldnl\nl
\newcommand{\nonl}{\renewcommand{\nl}{\let\nl\oldnl}}

\SetCommentSty{mycommfont}

\newcommand{\tech}{\text{{SpiNeMap}}}{}
\newcommand{\techcluster}{\text{{SpiNeCluster}}}{}
\newcommand{\techplacer}{\text{{SpiNePlacer}}}{}
\newcommand{\techhardware}{\text{{DynapSE}}}{}
\newcommand{\ours}{\text{{PSOPART}}}{}
\newcommand{\ineq}[1]{\footnotesize$#1$\normalsize}{}
\newcommand{\EnergyImprovement}{\text{{\ineq{\color{blue} 45\%}}}}{}
\newcommand{\LatencyImprovement}{\text{{\ineq{\color{blue} 21\%}}}}{}
\newcommand{\AccuracyImprovement}{\text{{\ineq{\color{blue} 12\%}}}}{}
\newcommand{\isiImprovement}{\text{{\ineq{\color{blue} 36\%}}}}{}
{}
\newcommand{\mr}[1]{\textcolor{black}{#1}}
\newcommand{\nr}[1]{\textcolor{black}{#1}}

\begin{document}
%
\title{\tech{}: Mapping Spiking Neural Networks on Neuromorphic Hardware}
\title{Mapping Spiking Neural Networks to Neuromorphic Hardware}
%
%

\author{Adarsha Balaji, Anup Das,
        Yuefeng Wu, Khanh Huynh, Francesco Dell'Anna, Giacomo Indiveri, \\Jeffrey L. Krichmar, Nikil Dutt, Siebren Schaafsma, and Francky Catthoor
}

%
%

\markboth{IEEE Transactions on Very Large Scale Integration (VLSI) Systems,~Vol.~XX, No.~X, Month~20XX}%
{Balaji \MakeLowercase{\textit{et al.}}: IEEE Transactions on Very Large Scale Integration (VLSI) Systems}
%



\maketitle

\begin{abstract}
\mr{Neuromorphic hardware platforms implement biological neurons and synapses to execute 
spiking neural networks (SNNs) in an energy-efficient manner. We present \tech{}, a design methodology to map SNNs to crossbar-based neuromorphic hardware, minimizing spike latency and energy consumption. \tech{} operates in {two} steps: \techcluster{} and \techplacer{}.
\techcluster{} is a heuristic-based  clustering technique to partition SNNs into clusters of synapses, where intra-cluster local synapses are mapped within crossbars of the hardware and inter-cluster global synapses are mapped to the shared interconnect. \techcluster{} minimizes the number of spikes on global synapses, which reduces spike congestion on the shared interconnect, improving application performance.
\techplacer{} then finds the best placement of local and global synapses on the hardware using a meta-heuristic-based approach
to minimize energy consumption and spike latency.}
We evaluate \tech{} using synthetic and realistic SNNs on the \techhardware{} neuromorphic hardware. We show that \tech{} reduces average energy consumption by \EnergyImprovement{} and average spike latency by \LatencyImprovement{}, compared to state-of-the-art techniques.
\end{abstract}

%
\IEEEpeerreviewmaketitle

\section{Introduction}\label{sec:Intro}

\IEEEPARstart{S}{piking} Neural Networks (SNNs) \cite{maass1997networks} are typically used for machine learning 
on energy-constrained devices \cite{das2018unsupervised,cao2015spiking,diehl2016conversion}. 
\mr{Neuromorphic platforms such as TrueNorth \cite{akopyan2015truenorth}, Loihi \cite{davies2018loihi}, and \techhardware{} \cite{Moradi_etal18} implement biological neurons and synapses, making them efficient in executing SNNs. Typically, these platforms consist of multiple crossbars with a shared time-multiplexed interconnect. A crossbar is a two-dimensional arrangement with $n$ rows, $n$ columns, and memory elements (to store synaptic weights) at every cross-point. 
Each crossbar can map at most $n$ synapses per neuron, meaning that a large SNN must be partitioned into synapses that map inside different crossbars (\textit{local synapses}) and those that map on the shared interconnect (\textit{global synapses}).
}

\mr{
A crossbar's size is usually kept small to reduce the energy consumed in driving high voltages through $n^2$ connections of a \ineq{n\times n} crossbar. For the \techhardware{} platform, with $n = 256$, a crossbar consumes 17pJ at 1.3V supply with SRAM-based synapses. This number is expected to reduce significantly when using non-volatile memory (NVM) synapses \cite{Burr2017NeuromorphicMemory}.
The shared interconnect in a neuromorphic hardware introduces spike congestion and latency to communicate spikes from one crossbar to another due to time-multiplexing, which impacts the inter-spike interval (ISI) \cite{sauer1995interspike}. This reduces application performance such as accuracy (see Section \ref{sec:motivation}).
}

\mr{
Many recent works demonstrate mapping of SNNs to a single crossbar \cite{ankit2018neuromorphic,zhang2018neuromorphic,xia2019memristive,lee2019system,wijesinghe2018all,wen2015eda}. In Section \ref{sec:results} we show how these techniques can be inefficient when applied to a multi-crossbar neuromorphic platform such as the \techhardware{}. There are only a few works that address SNN mapping to multi-crossbar neuromorphic hardware. These include the PACMAN \cite{galluppi2012hierachical}, NEUTRAMS \cite{ji2016neutrams}, and \ours{} \cite{das2018mapping}.
}

\mr{Compared to PACMAN and NEUTRAMS, which minimize crossbar usage, \ours{} {partitions} an SNN into local and global synapses, minimizing the number of spikes on the shared interconnect.
This optimization strategy reduces spike congestion and changes in ISI, which improves performance. \ours{} is designed for the \textit{shared bus} interconnect and it does not address the {placement} of local and global synapses to the neuromorphic hardware.
}

\mr{
Unfortunately, the shared bus becomes the latency and energy bottleneck for large SNNs, those with more than a million synapses \cite{orii2016advanced}. In recent years many new interconnects are explored for large-scale neuromorphic computing. Examples include the multi-stage networks-on-chip for the new TrueNorth platform \cite{DeBole2019TrueNorth:Years} and the segmented bus for the new \techhardware{} platform \cite{balajiexploring2019}.  For these new neuromorphic interconnects, the \ours{} technique has two limitations. First, the synapse partitioning approach is not scalable for large number of neurons and synapses. Second, different synapse placement strategies lead to different latency and energy consumption, which we show in Section \ref{sec:results}. Therefore, the placement problem can no longer be left unaddressed. 
}

We present \tech{}, a comprehensive design methodology to map SNNs to neuromorphic platforms\mr{, minimizing energy consumption and spike latency on the shared interconnect, and improving application performance. 
}

\noindent\textbf{Contributions} : Following are our novel contributions:
\begin{itemize}
	\item \textbf{\techcluster{}}: We propose a heuristic-based approach to partition an SNN into local and global synapses, reducing the number of spikes communicated on the shared interconnect.
	\item \textbf{\techplacer{}}: We propose  a meta-heuristic-based  approach to place local and global synapses on physical resources of a neuromorphic hardware, 
	reducing energy consumption and spike latency.
	\item We evaluate {\tech{}} on the \techhardware{} neuromorphic hardware using synthetic and realistic SNNs. 
	\item \mr{We evaluate different interconnect topologies and spike routing algorithms for emerging neuromorphic hardware.}
\end{itemize}

Table \ref{tab:contributions} compares our contributions against state-of-the-art techniques.
We evaluate \tech{} with SNN-based applications on the \techhardware{} hardware. We show that \tech{} reduces energy consumption by \EnergyImprovement{} and spike latency by \LatencyImprovement{} compared to state-of-the-art techniques. 

\begin{table}[t!]
	\renewcommand{\arraystretch}{1.2}
	\setlength{\tabcolsep}{2pt}
	\centering
	\begin{threeparttable}
	{\fontsize{7}{10}\selectfont
		\begin{tabular}{|c|c|c|p{3.8cm}|}
			\hline
			\textbf{Techniques} & \textbf{Partitioning} & \textbf{Placement} & \textbf{Optimization Objective}\\
			\hline
			\cite{ankit2018neuromorphic,zhang2018neuromorphic,xia2019memristive,lee2019system,wijesinghe2018all,wen2015eda} & $\times$ & $\times$ & Maximize single crossbar utilization\\
			NEUTRAMS \cite{ji2016neutrams} & $\surd$ & $\times$ & Minimize number of crossbars \\
			our \ours{} \cite{das2018mapping} & $\surd$ & $\times$ & Minimize spikes on global synapses\\
			\hline
			\hline
			\textcolor{blue}{\tech{}} & \textcolor{blue}{$\surd$} & \textcolor{blue}{$\surd$} & \textcolor{blue}{Minimize energy consumption and latency of neuromorphic hardware}\\
			\hline
	\end{tabular}}
	\begin{tablenotes}\footnotesize
        \item[$\surd$] Optimized by these approaches
        \item[$\times$] Not optimized by these approaches
    \end{tablenotes}
	\end{threeparttable}
	\caption{Contributions of \tech{} over the state-of-the-art approaches and our earlier work \cite{das2018mapping}.}
	\label{tab:contributions}
\end{table}

This paper is organized as follows. We provide background in Section \ref{sec:motivation}. We describe our design methodology of \tech{} in Section \ref{sec:method}. We present our evaluation setup in Section \ref{sec:evaluation} and results in Section \ref{sec:results}. We describe related works in Section \ref{sec:related_works}. We conclude the paper in Section \ref{sec:conslusion} with an outlook on the design of future neuromorphic platforms.

\section{\mr{Background}}
\label{sec:motivation}
\begin{figure}[t!]
	\centering
	\vspace{-10pt}
	\centerline{\includegraphics[width=0.99\columnwidth]{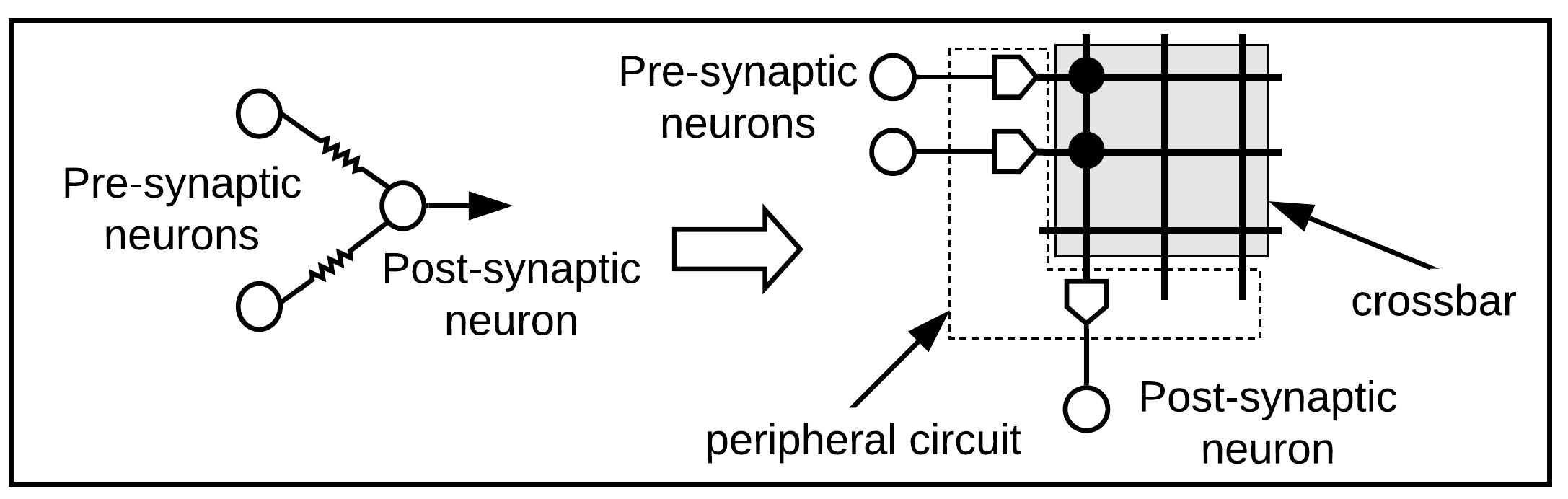}}
	\caption{Overview of how SNNs are mapped to a crossbar in a neuromorphic hardware.}
	\vspace{-10pt}
	\label{fig:overview_pcm}
\end{figure}

\mr{
Figure \ref{fig:overview_pcm} illustrates how a small SNN with two pre-synaptic neurons connected to a post-synaptic neuron is mapped to a crossbar. 
Spikes from a pre-synaptic neuron injects current into the crossbar, which is the product of spike voltage applied (i.e., input activation $x_i$) along the row with the conductance of the synaptic element at the cross-point (i.e., synaptic weight $w_{ij}$) following Ohm's law.
Current summations along columns are performed in parallel following Kirchhoff\textquotesingle s current law, and implement the sums $\sum_j w_{ij}x_i$, needed for forward propagation of neuron excitation $x_i$.
Beyond this supervised approach, recent works \cite{burr2017neuromorphic} have also developed peripheral structures necessary to implement online synaptic updates such as spike timing dependent plasticity (STDP) \cite{Rao2001Spike-timing-dependentLearning}.
}



\mr{
We demonstrate our design methodology for supervised machine learning approaches,
where an SNN is first trained with examples from the field and then deployed for inference with in-field data.
Performance 
is measured using \textit{accuracy}, which is assessed using inter-spike intervals (ISIs) \cite{phillips1991separate}.
}

\mr{
To define ISI, we consider an SNN with $N$ neurons and $S$ synapses, which is excited with an input over some finite interval of time $[0,T]$. Neural activities in this time interval generate $K$ spikes.
We organize these $K$ spikes based on their generation time and the source neuron of the SNN as
\begin{equation}
    \label{eq:spike_time}
    \footnotesize  \{t_1^1,t_2^1,\cdots,t_{k_1}^1\},\{t_1^2,t_2^2,\cdots,t_{k_2}^2\},\cdots,\{t_1^N,t_2^N,\cdots,t_{k_N}^N\},
\end{equation}
where $t_i^n$ is the time of the $i^\text{th}$ spike generated by the $n^\text{th}$ neuron in the time interval $[0,T]$ and $K = \sum_{i=1}^N k_i$. The ISI of this spike train is given by \cite{grun2010analysis}
\begin{equation}
    \label{eq:isi}
    \footnotesize I_i^n = t_i^n - t_{i-1}^n
\end{equation}
}

\vspace{-10pt}

\mr{
For a feedforward architecture \cite{LeCun2015DeepLearning}, (spiking) neurons are organized into layers, with one input layer, one or more hidden layers, and one output layer. For these architectures, accuracy is assessed from ISI of neurons in the (output) decision layer. For other architectures such as the Liquid State Machine (LSM) \cite{maass2002real}, ISI of critical neurons contribute to the accuracy.
}

\vspace{-10pt}
\mr{
Using CARLsim \cite{Chou2018CARLsim4} we can simulate different machine learning approaches and neural architectures, and extract ISI from any neuron in the architecture. This makes CARLsim our ideal starting point.
However, CARLsim is an application-level simulator meaning that hardware latencies are not incorporated. 
In a realistic scenario, ISI will be affected due to hardware latency arising from two sources -- 1) the \emph{fixed} latency within a crossbar to propagate current through synaptic elements and 2) the \emph{variable} latency of time multiplexing in the shared interconnect. In Section \ref{sec:method} we describe our framework \tech{} to obtain these latencies, starting from the application-level simulation results using CARLsim. 
}

\mr{
To incorporate hardware latency in ISI computation, Equation \ref{eq:spike_time} needs to be represented considering spike times at individual synapse-level. This is because different synapses have different latencies in neuromorphic hardware based on whether they are mapped within crossbars (i.e., local synapses) or on the shared interconnect (i.e., global synapses).
}

\mr{
The spike times on synapses are
\begin{equation}
    \label{eq:spike_time_synapse}
    \footnotesize \{\tau_1^1,\tau_2^1,\cdots,\tau_{k_1}^1\},\{\tau_1^2,\tau_2^2,\cdots,\tau_{k_2}^2\},\cdots,\{\tau_1^S,\tau_2^S,\cdots,\tau_{k_S}^S\}, 
\end{equation}
where $\tau_j^s$ is the $j^\text{th}$ spike on $s^\text{th}$ synapse and spike timings in the set \ineq{\{\tau_j^s\}} are obtained from spike timings in the set \ineq{\{t_i^n\}}.
We can similarly define ISI for this spike as
\begin{equation}
    \label{eq:isi_synapse}
    \footnotesize I_j^s = \tau_j^s - \tau_{j-1}^s
\end{equation}
We use the notation $\delta_j^s$ to represent the latency of the $j^\text{th}$ spike on $s^\text{th}$ synapse. The new ISI due to these latencies is
\begin{equation}
    \label{eq:isi_synapse_new}
    \footnotesize I_j^s|_\text{new} = \tau_j^s + \delta_j^s - \tau_{j-1}^s -  \delta_{j-1}^s
\end{equation}
The change in ISI (called \emph{ISI distortion}) is given by
\begin{equation}
    \label{eq:isi_distortion}
    \footnotesize I_j^s|_\text{distortion} = I_j^s|_\text{new} - I_j^s = \delta_j^s - \delta_{j-1}^s
\end{equation}
For local synapses, which are mapped within crossbars, all spikes have the same latency, i.e., \ineq{\delta_j^s = \delta_{j-1}^s}. So, the ISI distortion is \textit{zero}. For global synapses, different spikes of the same synapse can have different latencies due to the varying congestion and routing paths on the shared interconnect. These are the synapses that contribute to ISI distortion, i.e.,
\begin{equation}
    \label{eq:isi_distortion_final}
    \footnotesize I_j^s|_\text{distortion} =
    \begin{cases}
    0 & \text{if $s$ is mapped inside a crossbar}\\
    \delta_j^s - \delta_{j-1}^s & \text{if $s$ is mapped on the shared interconnect}
    \end{cases}
\end{equation}
}

\vspace{-10pt}

\mr{
ISI distortion due to the interconnect latency can lead to unacceptable accuracy loss.
Existing techniques \cite{ankit2018neuromorphic,zhang2018neuromorphic,xia2019memristive,lee2019system,wijesinghe2018all} minimize the latency inside crossbar, leaving the optimization of the interconnect latency to system designers.
In this work, we reduce the \textit{average ISI distortion} of spikes on all global synapses. Our framework can also perform other optimizations such as minimizing the \textit{maximum ISI distortion}.
}

\mr{
As we can clearly see from Equation \ref{eq:isi_distortion_final}, ISI distortion is due to the latency to time-multiplex spikes on the shared interconnect. This latency depends on the number of spikes that must be communicated via the shared interconnect at any given time (i.e., \textit{spike congestion}). Therefore, by reducing the number of spikes on global synapses we can reduce spike congestion, which would reduce ISI distortion and improve application performance. This is precisely the intuition behind our optimization strategy for the partitioning approach in our prior work \ours{} \cite{das2018mapping} and this current work. \nr{The difference is that the partitioning approach in this work is scalable to larger problem sizes than \ours{} (see Section \ref{sec:scalability} for comparison with \ours{}).}
}

\section{\tech{}: Mapping Spiking Neural Networks to Neuromorphic Hardware}
\label{sec:method}

\subsection{High-Level overview and difference with state-of-the-art}
\begin{figure}[t!]%
    \centering
    \subfloat[state-of-the-art approaches, e.g., NEUTRAMS \cite{ji2016neutrams}]{{\includegraphics[width=9cm]{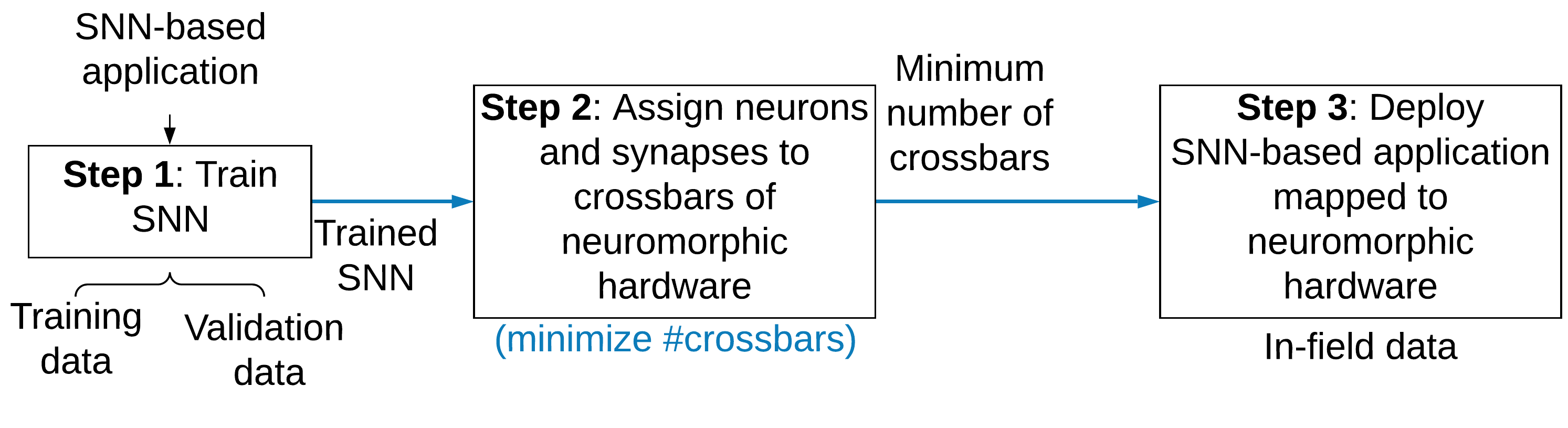} }}%
    \qquad
    \subfloat[our previous approach \ours{}\cite{das2018mapping}]{{\includegraphics[width=9cm]{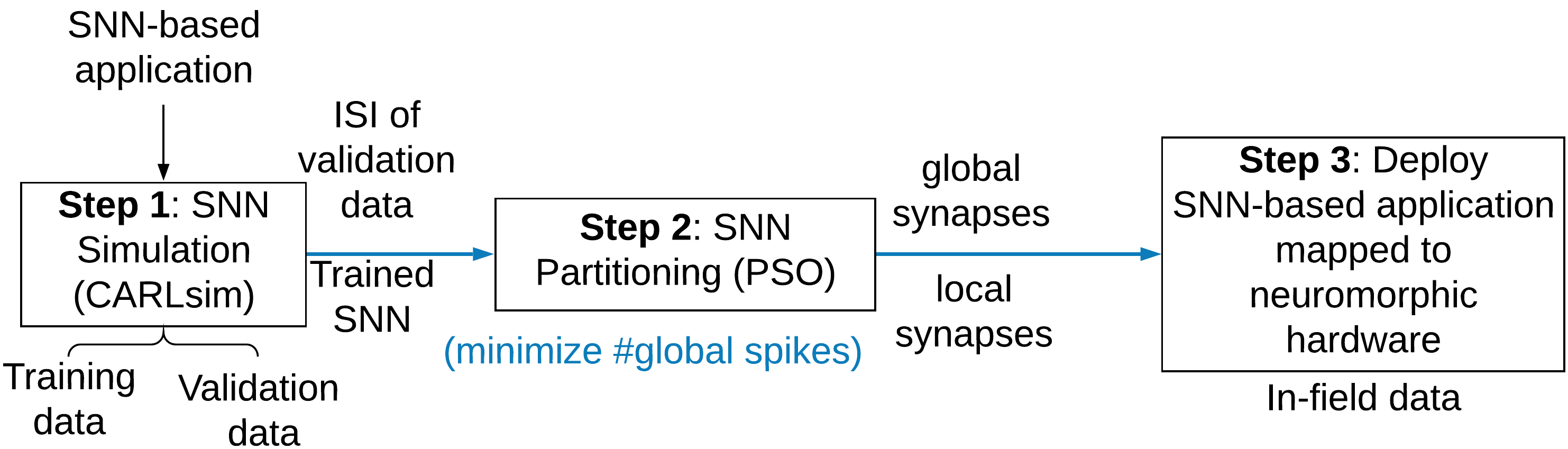} }}%
    \qquad
    \subfloat[our proposed design methodology \tech{}]{{\includegraphics[width=9cm]{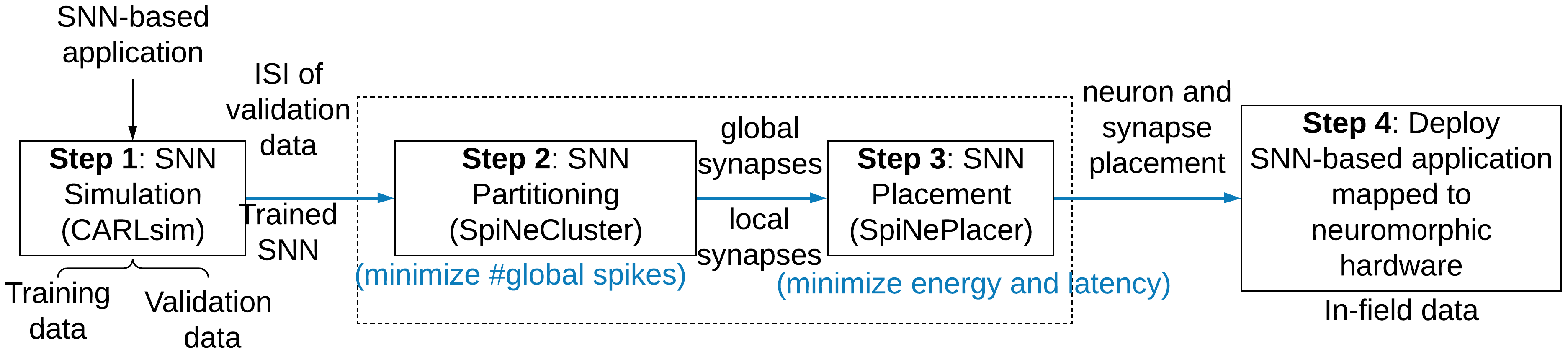} }}%
    \caption{A high-level overview of our \tech{} mechanism and its difference with state-of-the-art.}%
    \label{fig:overview}%
\end{figure}

In Figure \ref{fig:overview}, we illustrate 
our \tech{} methodology and its differences with state-of-the-art. In Figure \ref{fig:overview}(a), we illustrate how NEUTRAMS \cite{ji2016neutrams} and PACMAN \cite{galluppi2012hierachical} can be used to deploy SNN-based application on neuromorphic hardware. \mr{These approaches use 3 steps: \textit{Step 1)} train the SNN using training data and validate the trained model, \textit{Step 2)} pack neurons and synapses to crossbars, minimizing the resource requirements, and \textit{Step 3)} deploy the trained SNN mapped to the neuromorphic hardware for inference with in-field data.}

\mr{
Our previously-proposed \ours{} \cite{das2018mapping} also uses 3 steps to deploy SNN-based applications for inference (see Figure \ref{fig:overview}(b)). The difference in our prior approach is that we minimize the number of spikes on the global synapses to reduce ISI distortion, which improves application performance. To do so, we extract the spike count on every synapse of the SNN corresponding to the validation data used in SNN simulation. The spike count information is used by \ours{} to partition the SNN into local and global synapses using an instance of the particle swarm optimization (PSO) \cite{eberhart1995new}.
}

\mr{
In Figure \ref{fig:overview}(c), we illustrate our \tech{}. The key difference with our previously-proposed \ours{} is that we propose a 4-step methodology, with the new \textit{SNN Placement} step explicitly minimizing energy consumption and latency on the shared interconnect. This step is necessary for SNN mapping to large neuromorphic architectures with many crossbars. To do so, we extract not only the spike count on different synapses, but also their precise timing information by simulating the SNN in CARLsim. These information about spikes, also called \textit{spike trace}, are then used in \techplacer{} to simulate the exact latency and energy consumption, considering spike traffic on the shared interconnect. Overall, the \textit{SNN Partitioning} and \textit{Placement} steps jointly improve application performance, energy consumption, and spike latency.
}

\vspace{-10pt}
\subsection{Detailed design of SNN Partitioning via \techcluster{}}
\begin{figure}[t!]
	\centering
	\centerline{\includegraphics[width=0.6\columnwidth]{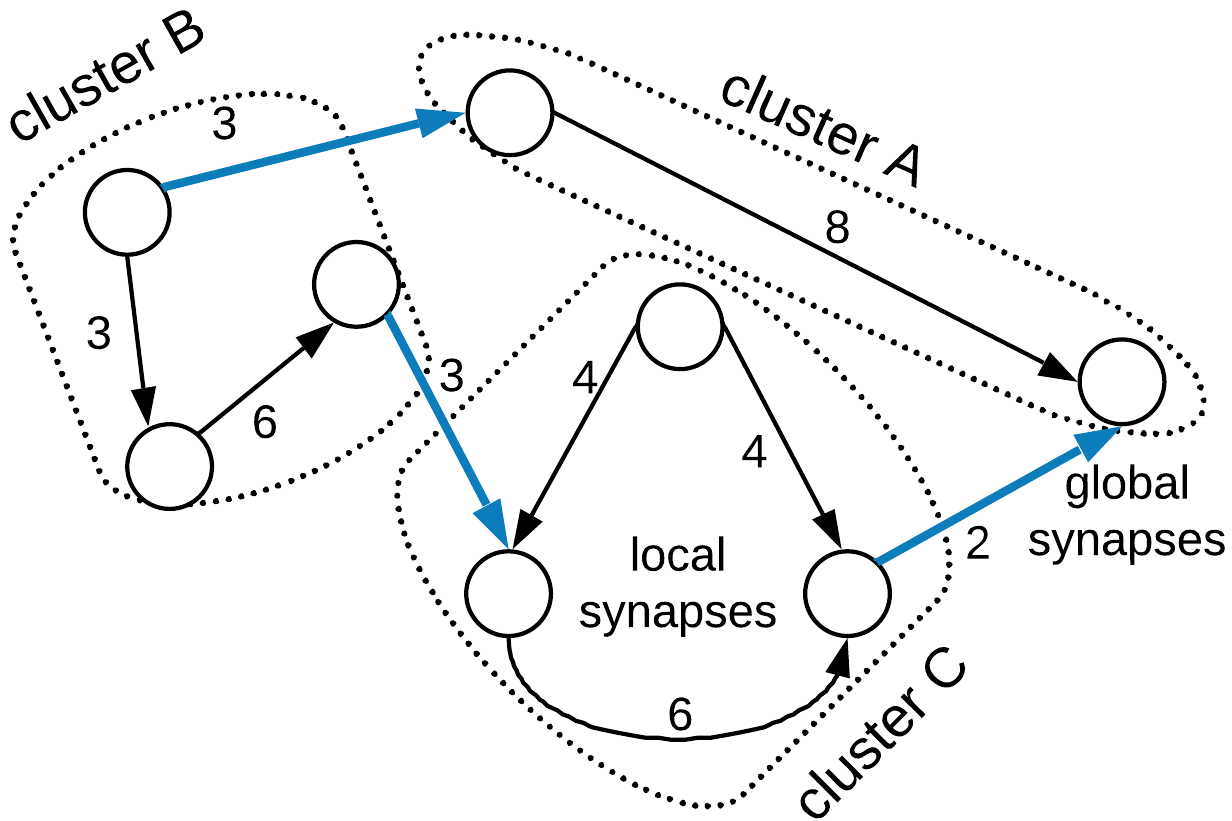}}
	\caption{An example illustrating how an SNN with 8 neurons is partitioned into 3 clusters with local and global synapses.}
	\label{fig:cluster_demo}
\end{figure}

In Figure \ref{fig:cluster_demo}, we illustrate an SNN partitioned into three clusters A, B, and C. The number of spikes communicated between a pair of neurons is indicated on its synapse. We also indicate the local synapses in black and the global ones in blue in this figure. In this example, the total number of spikes on global synapses is 8. To understand how \techcluster{} partitions an SNN,
we introduce the following notations.

\begin{algorithm}[t]
	\scriptsize{
		\ForEach{$C_i,C_j\in\mathcal{C}$}{
			\tcc{iterate over all cluster pairs}
			\tcc{begin 2-part procedure}
			$gs = $ total spikes between $C_i$ and $C_j$\;
			\While{True}{
				\ForEach{$n_i \in C_i$ and $n_j\in C_j$}{
					\If{$n_i$ and $n_j$ are not previously selected}{
						\textbf{Move} $n_i$ to $C_j$ and calculate $gs_1$\;
						\textbf{Move} $n_j$ to $C_i$ and calculate $gs_2$\;
						\textbf{Swap} $n_i$ and $n_j$ and calculate $gs_3$\;
						\textbf{Select} the option which lowers $gs$\;
						\textbf{Return} new partitions $C'_i,C'_j$\;
					}
				}
				$gs' = $ total spikes between $C'_i$ and $C'_j$\;
				\If{$gs' < gs$}{$gs = gs'$ and \textbf{break}\;}
			}
		\tcc{end 2-part procedure}
		}
	}
	\caption{SNN Clustering algorithm.}
	\label{alg:two_part_full}
\end{algorithm}

Let $\mathcal{G(N,S)}$ be an SNN with a set $\mathcal{N}$ of neurons, and a set $\mathcal{S}$ of synapses. A synapse $s_{i,j}$ connects neuron $n_i$ with $n_j$, and communicates $w_{i,j}$ spikes. 
\mr{Our objective is to partition this SNN into $k$ clusters.} 
Let $\mathcal{H(C,E)}$ be the partitioned SNN with a set $\mathcal{C}$ of clusters, and a set $\mathcal{E}$ of global synapses. This problem of transforming  $\mathcal{G(N,S)}\rightarrow\mathcal{H(C,E)}$ is a classic graph partitioning problem \cite{kernighan1970efficient}, and has been applied in many context, including task mapping on multiprocessor systems \cite{das2014communication}. The graph partitioning problem is already NP-complete \cite{garey1974some}, so heuristics are typically used to solve them \cite{fiduccia1982linear}. 
In our earlier work \ours{} \cite{das2018mapping}, we use an instance of particle-swarm optimization (PSO) \cite{kennedy2010particle} to solve this problem. However, the approach becomes \textit{intractable} as the size of the SNN increases. \mr{Here we propose a greedy approach, roughly based on the Kernighan-Lin Graph Partitioning algorithm \cite{kernighan1970efficient}, which we show to be scalable to large SNNs.}


We set 
$k = \ceil{\frac{|\mathcal{N}|}{n_c}}$, where $n_c$ is the number of neurons that can be accommodated per crossbar. \mr{We make this choice because \textit{by utilizing the minimum number of crossbars, the overall energy consumption of the hardware can be minimized} \cite{ankit2018neuromorphic,zhang2018neuromorphic,xia2019memristive,lee2019system,wijesinghe2018all}.} 
Next, we evenly (and arbitrarily) distribute neurons to these $k$ clusters. 
\mr{Starting from this arbitrary assignment, we analyze the change in the number of spikes on global synapses by moving a single neuron from one cluster to another, tracking and enforcing those changes that lead to minimum number of spikes on global synapses.}

We formalize these steps in Algorithm \ref{alg:two_part_full}. 
The algorithm applies a 2-part procedure (lines 2-17) to every cluster pair (with a total of $k \choose 2$ iterations). In the 2-part procedure, we first calculate the total number of inter-cluster spike ($gs$) with the two clusters (line 2).
Next, we select a pair of neurons $n_i$ and $n_j$ from the two selected clusters $C_i$ and $C_j$, respectively, such that neither $n_i$ nor $n_j$ is selected in the previous iterations (lines 4-5). We then perform three operations: (1) move $n_i\in C_i$ to cluster $C_j$ (if $C_j$ can accommodate more neurons) (line 6), (2) move $n_j\in C_j$ to cluster $C_i$ (if $C_i$ can accommodate more neurons) (line 7), and (3) swap $n_i$ and $n_j$ (line 8). We calculate the number of inter-cluster spike for each of these operations, and select the option that generates the maximum reduction of inter-cluster spike  compared to $gs$ (line 9). We return the new clusters (line 10). We  repeat the procedure (lines 4-13) while the number of inter-cluster spike continues to be reduced (lines 14-16).

\subsubsection{\underline{\mr{Time complexity}}} \mr{We compute the time complexity of Algorithm \ref{alg:two_part_full} as follows: Line 2-17 are executed \ineq{k \choose 2} times. At each iteration, lines 4-16 is iterated for every neurons of the a cluster with each cluster accommodating a maximum of \ineq{n_c} neurons. This time complexity is therefore
\begin{equation}
    \label{eq:time_complexity_1}
    \footnotesize \text{time complexity} = O\left({k\choose 2}\times n_c * n_c\right) = O\left(k^2\times n_c^2\right) = O\left(|\mathcal{N}|^2\right)
\end{equation}
where \ineq{k = \ceil{\frac{|\mathcal{N}|}{n_c}}}.
}

\begin{figure}[t!]
	\centering
	\centerline{\includegraphics[width=0.99\columnwidth]{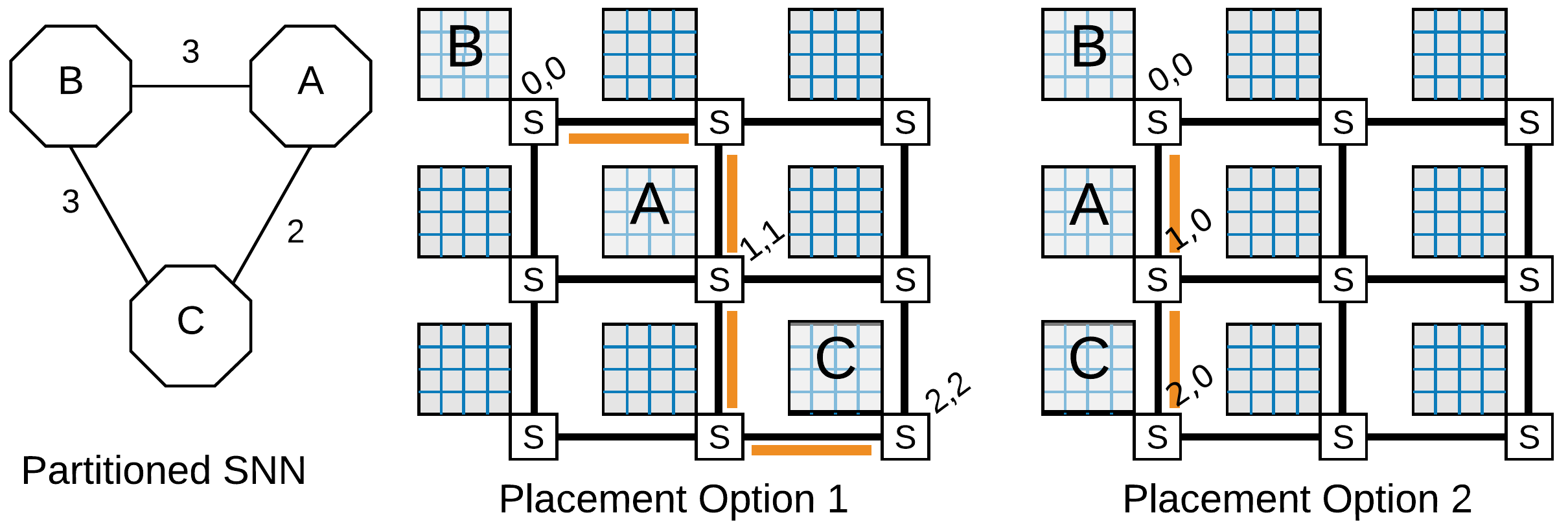}}
	\caption{Illustrating the impact of different placements of clusters of a partitioned SNN on a neuromorphic hardware.}
	\label{fig:placement}
\end{figure}

\begin{figure}[t!]
	\centering
	\centerline{\includegraphics[width=0.99\columnwidth]{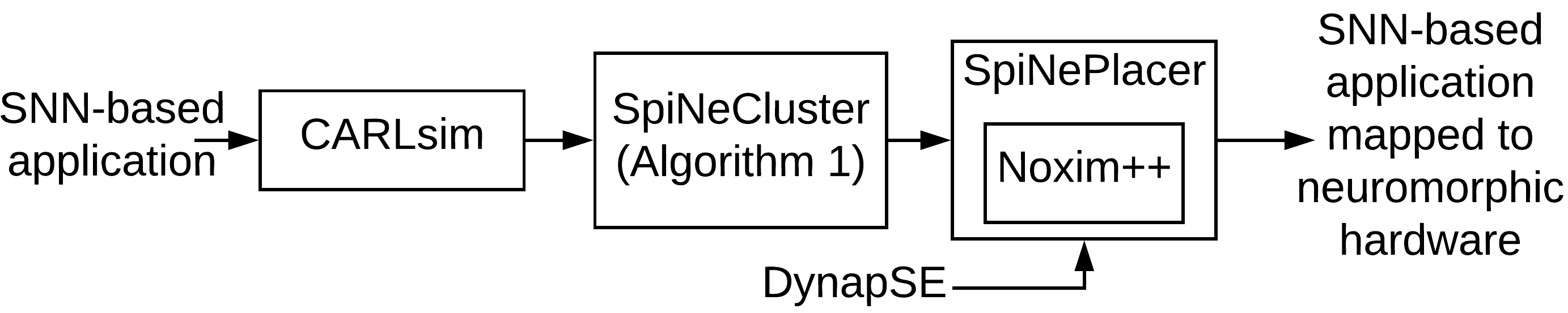}}
	\caption{Our design methodology \tech{}.}
	\label{fig:tech}
\end{figure}


\subsection{Detailed design of SNN Placement via \techplacer{}} 
\mr{
In Figure \ref{fig:placement}, we illustrate how a partitioned SNN (obtained via \techcluster{}) can be placed on to the hardware, which consists of three crossbars arranged in a mesh topology. Different placement alternatives lead to different interconnect lengths traversed by spikes to reach their destinations. This impacts both energy consumption and latency, meaning that the placement problem is no longer a trivial one, especially for large neuromorphic architectures (a limitation of NEUTRAMS \cite{ji2016neutrams}, PACMAN \cite{galluppi2012hierachical}, Eyeriss \cite{chen2017eyeriss}, and \ours{} \cite{das2018mapping}). 
}

\mr{
To accurately estimate the energy and latency impact of different placement alternatives, we have extended the Noxim \cite{catania2015noxim}, a cycle-accurate interconnect simulator, to support 1) simulation of \textit{spike traces} from CARLsim containing information (generation time, source neuron, and destination neuron) of every spike in an SNN and clustered to generate information about communication on global synapses, i.e., broadcast, multicast and one-to-one, 2) simulation of current and emerging interconnect topologies of neuromorphic architectures, 3) simulation of different routing algorithms, and 4) technology-specific energy and latency of interconnect wires and switches. We call our new framework \emph{Noxim++}.
}
\mr{
In Figure \ref{fig:tech} we present our design methodology \tech{}, illustrating how Noxim++ is integrated in the \techplacer{} and configured to model the \techhardware{} neuromorphic platform \cite{Moradi_etal18}. We now formalize the optimization problem of our \techplacer{}.
}

\mr{
We consider the mapping of a clustered SNN $\mathcal{H(C,E)}$ to the neuromorphic architecture $\mathcal{A(V,I)}$, where $\mathcal{V}$ is the set of crossbars in the architecture and $\mathcal{I}$ is the set of connections of these crossbars for a given interconnect topology.
}

\mr{
Mapping \ineq{M:\mathcal{H(C,E)}\rightarrow \mathcal{A(V,I)}} is represented by a logical matrix \ineq{(m_{ij}) \in \{0,1\}^{|\mathcal{C}|\times|\mathcal{V}|}}, where \ineq{m_{ij}} is defined as
\begin{equation}
    \label{eq:mapping_rep}
    \footnotesize m_{ij} = \begin{cases}
    1 & \text{if cluster } {c}_i \in \mathcal{C} \text{ is mapped to crossbar } {v}_j\in\mathcal{V}\\
    0 & \text{otherwise}
    \end{cases}
\end{equation}
}

\mr{
The constraints in this formulation are the following:
\begin{itemize}
    \item[1.] A cluster can be mapped to only one crossbar, i.e.,
    \begin{equation}
        \label{eq:mapping_constraint_1}
        \footnotesize \sum_j m_{ij} = 1~~~\forall i
    \end{equation}
    \item[2.] A crossbar can accommodate at most one cluster, i.e.,
    \begin{equation}
        \label{eq:mapping_constraint_2}
        \footnotesize \sum_i m_{ij} \leq 1~~~\forall j
    \end{equation}
\end{itemize}
}

\mr{
We use our Noxim++ framework to evaluate a mapping in terms of the optimization objective of \techplacer{}, i.e., to minimize energy consumption and spike latency on the interconnect. In has been shown in many prior works such as \cite{lee2007chip} that minimizing these metrics is equivalent to minimizing the \textit{average number of hops} that spikes communicate before reaching their destination (see also the formulations in Section \ref{sec:des_metrics}). 
Let \ineq{\mathcal{L}_i} be the average hop count for the cluster mapping \ineq{M_i} obtained using Noxim++, i.e., \ineq{\mathcal{L}_i = \text{Noxim++}(M_i)}.
The optimization objective of our \techplacer{} is
to find the mapping with the minimum average hop count, i.e.,
\begin{equation}
    \label{eq:opt_obj}
    \footnotesize \mathcal{L}_\text{min} = \mathcal{L}_\text{a}, \text{ where } a = \argmin\{\text{Noxim++}(M_i) | i\in 1,2,\cdots,N_m\},
\end{equation}
where $N_m$ is the total number of mappings evaluated. Of different techniques to generate and evaluate cluster mappings, we use an instance of PSO, which we describe next.
}

 In general, PSO finds the optimum solution to a fitness function $F$. \mr{There can be several particles in the swarm. The position of these particles are solutions to the fitness functions, and they represent cluster mappings, i.e., \ineq{M}'s in Equation \ref{eq:opt_obj}.} Each particle also has a velocity with which it moves in the search space to find the optimum solution. During the movement, a particle updates its position and velocity according to its own experience (closeness to the optimum) and also experience of its neighbors. We introduce the following notations for PSO.
 
\begin{footnotesize}
 	\begin{align}
 	\label{eq:pso_defn}
 	D &= \text{dimensions of the search space}\\
 	n_p &= \text{number of particles in the swarm}\nonumber \\
 	\mathbf{\Theta} = \{\mathbf{\theta}_l\in\mathbb{R}^{D}\}_{l=0}^{n_p-1} &= \text{positions of particles in the swarm}\nonumber \\
 	\mathbf{V} = \{\mathbf{v}_l\in\mathbb{R}^{D}\}_{l=0}^{n_p-1} &= \text{velocity of particles in the swarm}\nonumber 
 	\end{align}
 \end{footnotesize}

\mr{Here \ineq{\theta_l} is the position of the \ineq{l^\text{th}} particle in the swarm, and translates to the mapping \ineq{M_l}. \ineq{D} is therefore the dimension of the logical mapping matrix \ineq{M}, i.e., \ineq{D = |\mathcal{C}|\times|\mathcal{V}|}}.

Position and velocity updates are performed according to the following equation.

\begin{footnotesize}
	\begin{align}
	\label{eq:pos_vel_update}
	\mathbf{\Theta}(t+1) &= \mathbf{\Theta}(t) + \mathbf{V}(t+1)\\
	\mathbf{V}(t+1) &= \mathbf{V}(t) + \varphi_1\cdot\Big(P_{\text{best}}-\mathbf{\Theta}(t)\Big) + \varphi_2\cdot\Big(G_{\text{best}}-\mathbf{\Theta}(t)\Big)\nonumber
	\end{align}
\end{footnotesize}
\normalsize where $t$ is the iteration number, $\varphi_1,\varphi_2$ are constants and $P_{\text{best}}$ (and $G_{\text{best}}$) is the particles own (and neighbors) experience. 
\mr{
The fitness function is then
\begin{equation}
    \label{eq:fitness}
    \footnotesize F(\theta_l) = \mathcal{L}_l = \text{Noxim++}(M_l)
\end{equation}
}

\mr{Once the fitness function is computed for all particles in the \textit{swarm}, the personal best position of each particle (\ineq{P_\text{best}^t}) and the global best position of the swarm (\ineq{G_\text{best}}) are updated using Equation \ref{eq:pbest}.
}

\mr{
\begin{footnotesize}
    \begin{align}
    \label{eq:pbest}
        && P_\text{best}^t = F(\theta_t) \text{ if } F(\theta_t) < F(P_\text{best}^t)\nonumber \\
        && G_\text{best} = \displaystyle \min_{t=0,\dots n_p-1} P_\text{best}^t
    \end{align}
\end{footnotesize}
}

Due to the binary formulation of the mapping problem (see Equation \ref{eq:mapping_rep}), we need to binarize the velocity and position of Equation \ref{eq:pso_defn}, which we illustrate below.

\begin{footnotesize}
	\begin{align}
	\label{eq:binarization}
	&\hat{\mathbf{V}} = \texttt{sigmoid}(\mathbf{V}) = \frac{1}{1+\texttt{e}^{-\mathbf{V}}} \nonumber \\
	&	\hat{\Theta} = \begin{cases}
	0 \text{~~if } \texttt{rand()} < \hat{\mathbf{V}}\\
	1 \text{~~otherwise }
	\end{cases}
	\end{align}
\end{footnotesize}


In finding a new position of a PSO particle, we use the two {constraints} in Equations \ref{eq:mapping_constraint_1} \& \ref{eq:mapping_constraint_2}.

\begin{figure}[t!]
	\centering
	\centerline{\includegraphics[width=0.99\columnwidth]{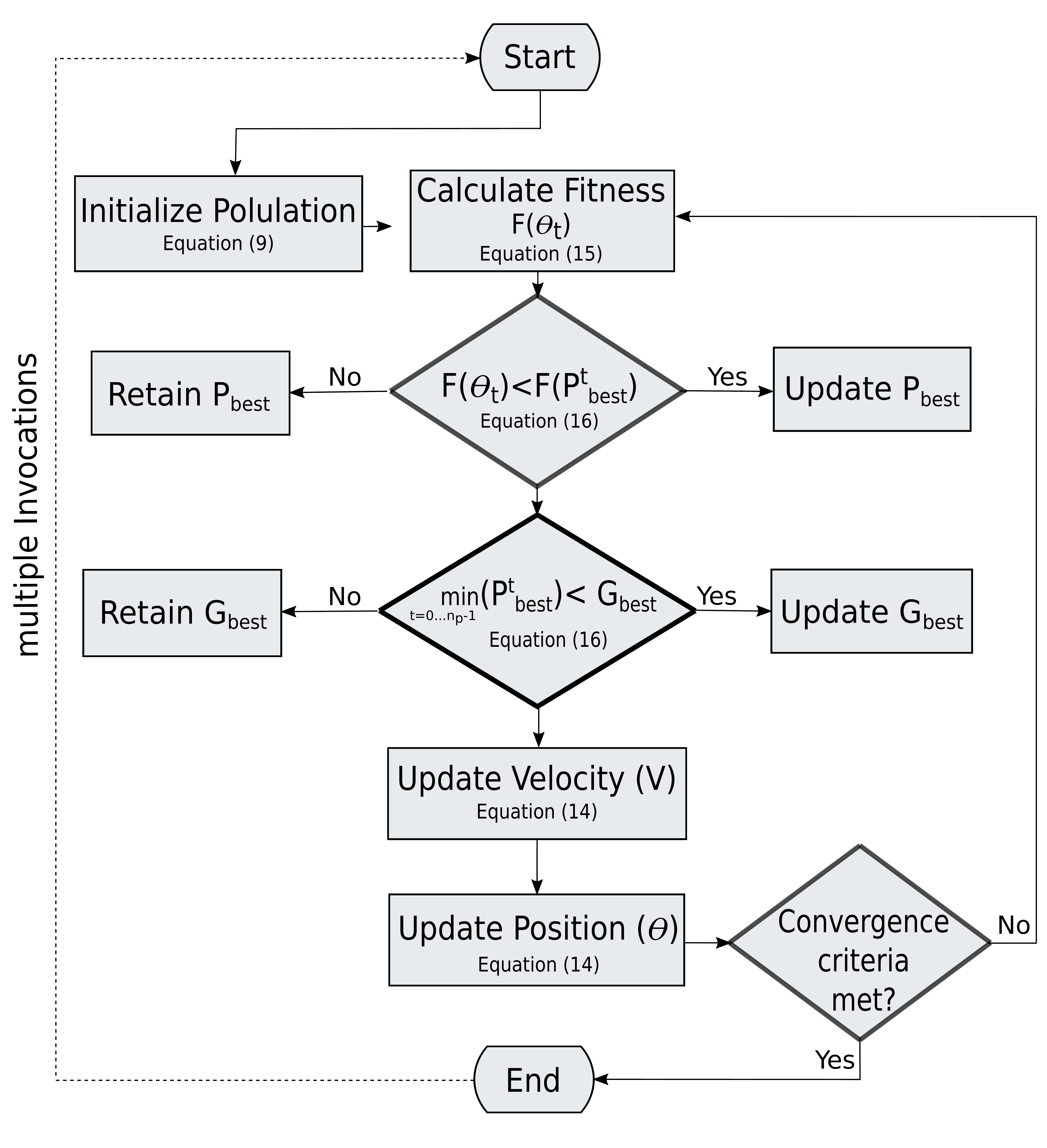}}
	\caption{Flow chart of our PSO algorithm.}
	\label{fig:pso_flow}
\end{figure}

\subsubsection{\underline{\mr{PSO Algorithm}}}
\mr{
In Figure \ref{fig:pso_flow}, we describe our iterative PSO algorithm that uses the analytical formulations we introduced in Equations \ref{eq:mapping_constraint_1}-\ref{eq:binarization}. The algorithm begins by initializing the position of the PSO particles that satisfies constraints \ref{eq:mapping_constraint_1} \& \ref{eq:mapping_constraint_2}. Then the algorithm runs for \ineq{n_\text{ISO}} iterations. At each iteration, the PSO algorithm evaluates the fitness function (Equation \ref{eq:fitness}) and updates its position based on the local and global best positions (Equation \ref{eq:pos_vel_update}), binarizing these updates using Equation \ref{eq:binarization}. The time complexity of the PSO algorithm is therefore \ineq{O(n_\text{ISO}\times \text{operations in each iteration})}, where operations in each iteration is proportional to the PSO dimension \ineq{D = |\mathcal{C}|\times|\mathcal{V}|} and the number of particles \ineq{n_p}. We represent the time complexity as
\begin{equation}
    \label{eq:time_complexity_2}
    \footnotesize \text{time complexity of PSO} = O\left(n_\text{ISO}\times n_p\times |\mathcal{C}|\times|\mathcal{V}|\right)
\end{equation}
}




\subsection{Justification of \tech{}'s design choices}
In this section, we motivate \tech{}'s design choices. 

\subsubsection{\mr{\underline{Minimize spike count at the partitioning stage}}}
\label{sec:justify_spike_minimization}
\begin{figure}[t!]
	\centering
	\centerline{\includegraphics[width=0.99\columnwidth]{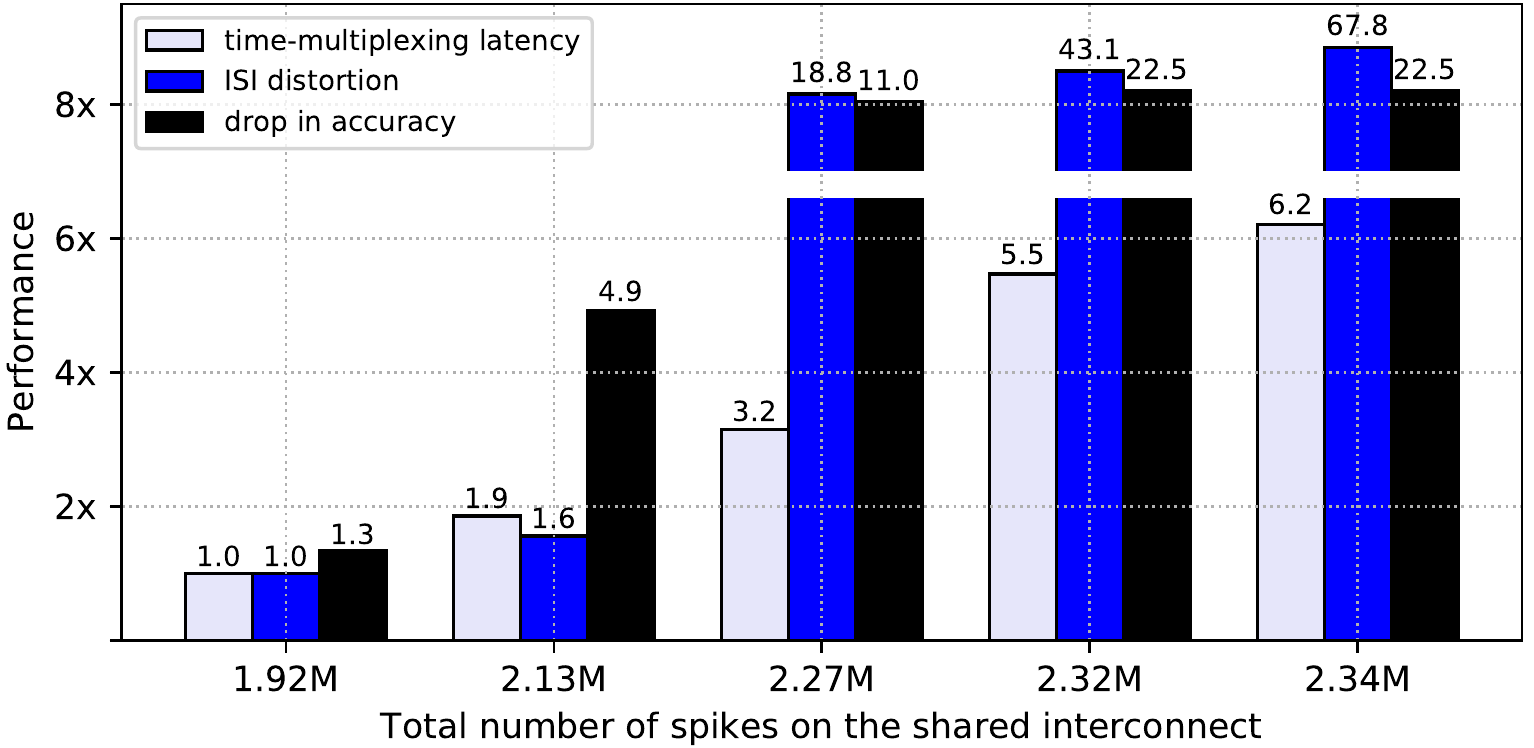}}
	\caption{Latency, ISI distortion, and accuracy as a function of the number of spikes on the global interconnect for the handwritten digit recognition example.}
	\label{fig:mnist_new_accuracy}
\end{figure}

\mr{To justify our optimization objective of minimizing the number of spikes at the partitioning stage of our design methodology, we conducted an experiment with the hand written digit recognition example, where as the number of spikes on the shared interconnect is increased, the latency and average ISI distortion on the time-multiplexed interconnect, and the classification accuracy are recorded. We use the hardware configuration of \techhardware{}, with four crossbars organized in a 2x2 mesh with XY routing algorithm. Each crossbar can accommodate 256 neurons. We report these results in Figure \ref{fig:mnist_new_accuracy}, with the latency and ISI distortion normalized to the case with minimum number of spikes on the shared interconnect. The drop in accuracy is calculated with respect to the accuracy obtained when the number of spikes on the shared interconnect in the minimum.}

\mr{We observe that as the number of spikes on the shared interconnect increases, the latency increases, increasing the ISI distortion. This lowers the application accuracy. We observe a similar behavior for all our evaluated applications.}

\subsubsection{\mr{\underline{Integration of Noxim++ within PSO}}}
\label{sec:justify_noxim_pso}
\mr{The average hop count of spikes communicated between clusters (i.e., crossbars) on the shared interconnect depends on 1) the cluster mapping \ineq{M} and 2) the routing algorithm that \textit{dynamically} routes spikes on the interconnect to avoid congestion of interconnect links. Our PSO incorporates cluster mapping in the fitness function. Due to the dynamic nature of spike routing for congestion avoidance, we need to simulate the cycle-accurate behavior of the interconnect for every mapping with the spike trace generated from CARLsim to accurately compute the hop distance that each spike traverses before reaching its destination. This motivates our strategy to integrate Noxim++ within PSO to minimize the average hop count.
}

\subsubsection{Using PSO only for \techplacer}
We use binary particle swarm optimization (PSO) \cite{eberhart1995new}, an evolutionary computing technique inspired by social behaviors such as bird flocking and fish schooling. Evolutionary computing techniques, in general, are efficient in avoiding being stuck at local optima. Additionally, PSO is computationally less expensive with faster convergence compared to its counterparts such as genetic algorithm (GA) or simulated annealing (SA).

In our earlier work \cite{das2018mapping}, we use PSO for SNN partitioning (equivalent of \techcluster{}).  In this work we use PSO only for \techplacer{} and a greedy approach for \techcluster{}. The rationale behind this is as follows.

Had PSO been used for \techcluster{}, the total number of dimensions for each particle in PSO would be $D = |\mathcal{N}|\times|\mathcal{C}|$. The total number of dimensions of each particle in the PSO of \techplacer{} is $D = |\mathcal{C}|\times|\mathcal{V}|$. In Table \ref{tab:pso_dimn}, we compare these dimensions for different SNN sizes, with a fixed neuromorphic hardware (16 crossbars, with 256 neurons each).

\begin{table}[h!]
	\setlength{\tabcolsep}{5pt}
	\renewcommand{\arraystretch}{1.3}
	\centering
	\begin{tabular}{c||c|c}
		\hline
		\textbf{\# of} & \multicolumn{2}{c}{\textbf{PSO dimensions ($D$) for}}\\ \cline{2-3}
		\textbf{SNN neurons} & \textbf{SNN partitioning} & \textbf{SNN placement}\\
		\hline
		1,000 & 16,000 & 64\\
		2,000 & 32,000 & 128\\
		3,000 & 48,000 & 192\\
		4,000 & 64,000 & 256\\
		\hline
	\end{tabular}
	\caption{Dimensions of PSO to solve partitioning and placement problems, for different SNN sizes on a fixed neuromorphic hardware with 16 crossbars, and 256 neurons each.}
	\label{tab:pso_dimn}
\end{table}

As we can clearly see from Table \ref{tab:pso_dimn}, the PSO problem of partitioning soon becomes intractable with modest size SNNs, even if we restrict to 1000 particles (each with dimensions $D$) in the swarm. To keep the solution time reasonable, we therefore, use PSO only for the placement problem (viz. \techplacer{}), and use a greedy approach instead for the partitioning problem (viz. \techcluster).

\section{Evaluation Methodology}
\label{sec:evaluation}
We build \tech{} with the following system components.
\begin{itemize}
	\item \textbf{CARLsim} \cite{Chou2018CARLsim4} : A GPU accelerated simulator used to train and test SNN-based applications. CARLsim reports spike times for every synapse in the SNN. 
	\item \textbf{Noxim++} \cite{catania2015noxim} : A trace-driven and cycle-accurate interconnect simulator for multiprocessor systems. We extend it (1) to incorporate crossbar-based neuromorphic hardware, (2) to communicate spikes (rather than data packets), and (3) to generate key performance statistics such as energy consumption, spike latency and ISI distortion. In our \tech{} mechanism, the spike timing information from CARLsim is used as trace and is input to Noxim++ to generate the performance statistics.
	\item  \textbf{\techhardware{}} \cite{Moradi_etal18}: We configure Noxim++ to model the \techhardware{} neuromorphic hardware at 65nm technology nodes with 256 neurons per crossbar. \mr{The crossbars are interconnected using a multi-stage networks-on-chip (NoCs) \cite{benini2002networks}.} We extract the latency and energy numbers of each crossbar from silicon data \cite{indiveri2015neuromorphic}. We also use the analytical performance and energy model of the interconnect network at 65nm technology. Finally, we use predictive technology mapping (PTM) \cite{zhao2006new} to scale technology parameters to 28nm nodes.
\end{itemize}

\subsection{Simulation environment}
We conduct all experiments on a system with 8 CPUs, 32GB RAM, and NVIDIA Tesla GPU, running Ubuntu 16.04. 

\subsection{Evaluated applications}
In order to evaluate the effectiveness of \tech{}, we use 7 synthetic and 8 realistic SNN applications. These applications are described in Table \ref{tab:apps}. We indicate the synthetic applications with the letter `S' followed by a number (e.g., S\_1000), where the number represents the total number of neurons in the synthetic SNN. We use 7 synthetic SNN applications with number of neurons between 1000 to 4000. In column 3 of this table, we indicate the number of synapses in the networks, while in column 4 we describe the corresponding SNN topology. The total number of synapse in these synthetic applications ranges from 240,000 in S\_100 to 3.75M in S4000.

\begin{table}[t!]
	\renewcommand{\arraystretch}{0.8}
	\setlength{\tabcolsep}{2pt}
	\centering
	\begin{threeparttable}
	{\fontsize{6}{10}\selectfont
		\begin{tabular}{cc|cl|c}
			\hline
			\textbf{Category} & \textbf{Applications} & \textbf{Synapses} & \textbf{Topology} & \textbf{Spikes}\\
			\hline
			\multirow{7}{*}{synthetic} & S\_1000 & 240,000 & FeedForward (400, 400, 100) & 5,948,200\\
			& S\_1500 & 300,000 & FeedForward (500, 500, 500) & 7,208,000\\
			& S\_2000 & 640,000 & FeedForward (800, 400, 800) & 45,807,200\\
			& S\_2500 & 1,440,000 & FeedForward (900, 900, 700) & 66,972,600\\
			& S\_3000 & 2,000,000 & FeedForward (1000, 1000, 1000) & 155,123,000\\
			& S\_3500 & 2,500,000 & FeedForward (1000, 1000, 1500) & 46,476,000\\
			& S\_4000 & 3,750,000 & FeedForward (1500, 1500, 1000) & 149,580,500\\
			\hline
			\multirow{8}{*}{realistic} & ImgSmooth \cite{Chou2018CARLsim4} & 136,314 & FeedForward (4096, 1024) & 17,600\\
			& EdgeDet \cite{Chou2018CARLsim4} & 272,628 &  FeedForward (4096, 1024, 1024, 1024) & 22,780\\
			& MLP-MNIST \cite{diehl2015unsupervised} & 79,400 & FeedForward (784, 100, 10) & 2,395,300\\
			& HeartEstm \cite{das2018unsupervised} & 636,578 & Recurrent & 3,002,223\\
			& HeartClass \cite{balaji2018power} & 2,396,521 & CNN\tnote{1} & 1,036,485\\
			& CNN-MNIST \cite{mlperf} & 159,553 & CNN\tnote{2} & 97,585\\
			& LeNet-MNIST \cite{mlperf} & 1,029,286 & CNN\tnote{3} & 165,997\\
			& LeNet-CIFAR \cite{mlperf} & 2,136,560 & CNN\tnote{4} & 589,953 \\
			\hline
	\end{tabular}}
	\begin{tablenotes}\scriptsize
        \item[1.] Input(82x82) - [Conv, Pool]*16 - [Conv, Pool]*16 - FC*256 - FC*6
        \item[2.] Input(24x24) - [Conv, Pool]*16 - FC*150 - FC*10
        \item[3.] Input(32x32) - [Conv, Pool]*6 - [Conv, Pool]*16 - Conv*120 - FC*84 - FC*10
        \item[4.] Input(32x32x3) - [Conv, Pool]*6 - [Conv, Pool]*6 - FC*84 - FC*10
    \end{tablenotes}
	\end{threeparttable}
	\caption{7 synthetic and 8 realistic applications we use to evaluate \tech{}.}
	\label{tab:apps}
\end{table}

\mr{We use eight realistic applications: \textit{image smoothing} (ImgSmooth) \cite{Chou2018CARLsim4} on 64x64 images, \textit{edge detection} (EdgeDet) \cite{Chou2018CARLsim4} on 64x64 images using difference-of-Gaussian, \textit{multi-layer perceptron (MLP)-based handwritten digit recognition} (MLP-MNIST) \cite{diehl2015unsupervised} on 28x28 images of handwritten digits, \textit{ECG-based heart-rate estimation} (HeartEstm) \cite{das2018unsupervised}, \textit{ECG-based heart-beat classification} (HeartClass) \cite{balaji2018power}, \textit{CNN-based digit classification} (CNN-MNIST) \cite{springenberg2014striving,mlperf}, \textit{CNN-based digit classification with LeNet} (LeNet-MNIST) \cite{mlperf}, and \textit{CNN-based CIFAR image classification with LeNet} (LeNet-CIFAR) \cite{mlperf}.
We note that the last three applications are part of the MLPerf benchmark suite \cite{mlperf} and developed using analog computation model. We convert these applications into spike-based model using the CNN-to-SNN conversion tool N2D2 \cite{n2d2}.
}

Finally, in the last column of Table \ref{tab:apps} we report the total number of spikes for these applications obtained through simulation of the representative validation data using CARLsim \cite{Chou2018CARLsim4}. 
\mr{
The spike trace from CARLsim is clusted using \techcluster{}, and placed on crossbars using \techplacer{}.
}

\subsection{Evaluated state-of-the-art techniques}
\mr{We evaluate the following four approaches.
\begin{itemize}
	\item \underline{Baseline}:  The Baseline uses NEUTRAMS \cite{ji2016neutrams} to cluster SNNs, minimizing the use of crossbars.
	\item \underline{SCO}:  The SCO approach uses the framework of \cite{lee2019system} to balance the utilization of crossbars in the hardware.
	\item \underline{\ours{}}: Our previously-proposed \ours{} \cite{das2018mapping} clusters SNNs to minimize the total number of spikes on the shared interconnect.
	\item \underline{\tech{}}: Our \tech{} uses (1) \techcluster{} to partition SNNs into clusters to minimize the total number of spikes on the shared interconnect and (2) \techplacer{} to optimize the placement of clusters to crossbars of the neuromorphic hardware to minimize energy consumption and latency on the shared interconnect.
\end{itemize}
}

\subsection{\mr{Evaluated metrics}}\label{sec:des_metrics}
\mr{We evaluate all four approaches in terms of the following metrics for every application.
\begin{itemize}
    \item \underline{Total number of spikes on the shared interconnect:} This is the number of spikes (\ineq{N_s}) on the shared interconnect obtained after mapping synapse clusters to crossbars of the neuromorphic hardware.
    \item \underline{Spike latency on the shared interconnect:} This is the delay experienced by spikes before reaching their destination, averaged over all spikes \cite{lee2007chip}, i.e.,
    \begin{equation}
        \label{eq:lat}
        \footnotesize L = \sum_{i=1}^{N_s} [(h_i - 1)*l_w + h_i*l_s] / N_s,
    \end{equation}
    where \ineq{h_i} is the number of hops that spikes traverses between the source crossbar and destination crossbar, \ineq{l_w} is the delay on the wires connecting two crossbars, and \ineq{l_s} is the delay of the hop.
    \item \underline{Energy consumption on the shared interconnect:} This is the total energy consumed by all spikes on the shared interconnect \cite{lee2007chip}, i.e.,
    \begin{equation}
        \label{eq:en}
        \footnotesize E = \sum_{i=1}^{N_s} [(h_i - 1)*e_w + h_i*e_s],
    \end{equation}
    where \ineq{e_w} and \ineq{e_s} are the energy consumption on the wires and hops, respectively.
    \item \underline{Average ISI distortion:} This is motivated in Section \ref{sec:motivation} and computed using Equation \ref{eq:isi_distortion_final}, averaged over all spikes, i.e.,
    \begin{equation}
        \label{eq:isid}
        \footnotesize I = \sum_{i=1}^{N_s} I_i|_{distortion} / N_s,
    \end{equation}
\end{itemize}
}
\vspace{-10pt}
\section{Results and Discussions}
\label{sec:results}

\subsection{Improvement summary}
\label{sec:summary}
In Table \ref{tab:compare_sota}, we summarize the average improvements of \tech{} \mr{against the Baseline \cite{ji2016neutrams}, SCO \cite{lee2019system}, and our previously-proposed \ours{}\cite{das2018mapping}.}

\begin{table}[t!]
	\renewcommand{\arraystretch}{1.0}
	\setlength{\tabcolsep}{1pt}
	\centering
	{\fontsize{8}{10}\selectfont
		\begin{tabular}{|l|c|c|c|c|}
			\hline
			\multirow{3}{*}{\textbf{\tech{}}} & \textbf{Energy} & \textbf{Spike} & \textbf{ISI} & \textbf{Application} \\
			& \textbf{Consumption} & \textbf{Latency} & \textbf{Distortion} & \textbf{Accuracy}\\
			& \textbf{(Sec. \ref{sec:energy_consumption})} & \textbf{(Sec. \ref{sec:latency})} & \textbf{(Sec. \ref{sec:isi})} & \textbf{(Sec. \ref{sec:app_accuracy})}\\
			\hline
			vs. Baseline \cite{ji2016neutrams} & \EnergyImprovement{} & \LatencyImprovement{} &\isiImprovement{} & \AccuracyImprovement{}\\
			vs. SCO \cite{lee2019system} & \ineq{40\%} & \ineq{27\%} & \ineq{39\%} & \ineq{20\%}\\
			vs. \ours{} \cite{das2018mapping} & \ineq{20\%} & \ineq{13\%} & \ineq{23\%} & \ineq{5\%}\\
			\hline
	\end{tabular}}
	\caption{\mr{Average improvement summary using \tech{} for all our evaluated applications.}}
	\label{tab:compare_sota}
\end{table}

We now describe these results in details.

\vspace{-10pt}
\subsection{Energy consumption on the shared interconnect}
\label{sec:energy_consumption}
\begin{figure}[h!]
	\centering
	\centerline{\includegraphics[width=0.99\columnwidth]{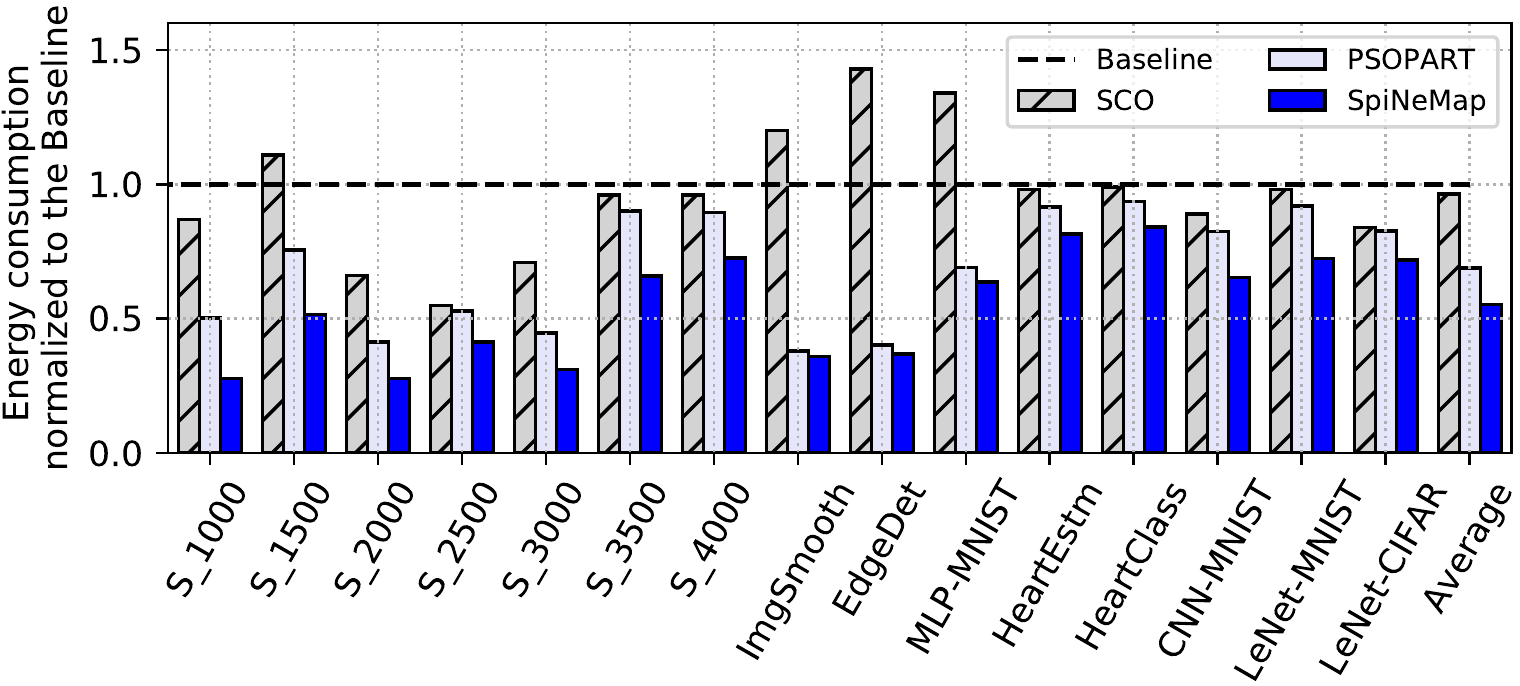}}
	\caption{Energy consumption on the shared interconnect normalized to the Baseline.}
	\label{fig:energy}
\end{figure}

In Figure \ref{fig:energy}, we report the energy consumption on the shared interconnect of each of our applications for each of our evaluated systems normalized to the Baseline. We make the following three observations.

\mr{\textit{First}, the average energy consumption of SCO is very similar to that of the Baseline. For some workloads such as S\_2500 it achieves 45\% lower energy consumption, while for other workloads such as EdgeDet it has 43\% higher energy consumption than the Baseline. These differences are due to the earlier discussed distinction in the optimization objective for these two approaches.} \textit{Second}, \ours{} has \ineq{31\%} lower average energy consumption than the Baseline. This reduction is because \ours{} minimizes the total number of global spikes, which also reduces the energy consumption on the shared interconnect.
\textit{Third}, \tech{} has the lowest energy consumption of all our evaluated systems (\EnergyImprovement{} lower average energy consumption than the Baseline, \ineq{40\%} lower than SCO, and \ineq{20\%} lower than \ours{}). 
\mr{
These improvements are because of \tech{}'s optimization policies: 1) \techcluster{} reduces the total number of spikes on the shared interconnect, which lowers energy consumption, and 2) \techplacer{} places the clusters on crossbars of the hardware to minimize both latency and energy consumption on the shared interconnect.
}

\subsection{Spike latency on the shared interconnect}
\label{sec:latency}
\begin{figure}[h!]
	\centering
	\centerline{\includegraphics[width=0.99\columnwidth]{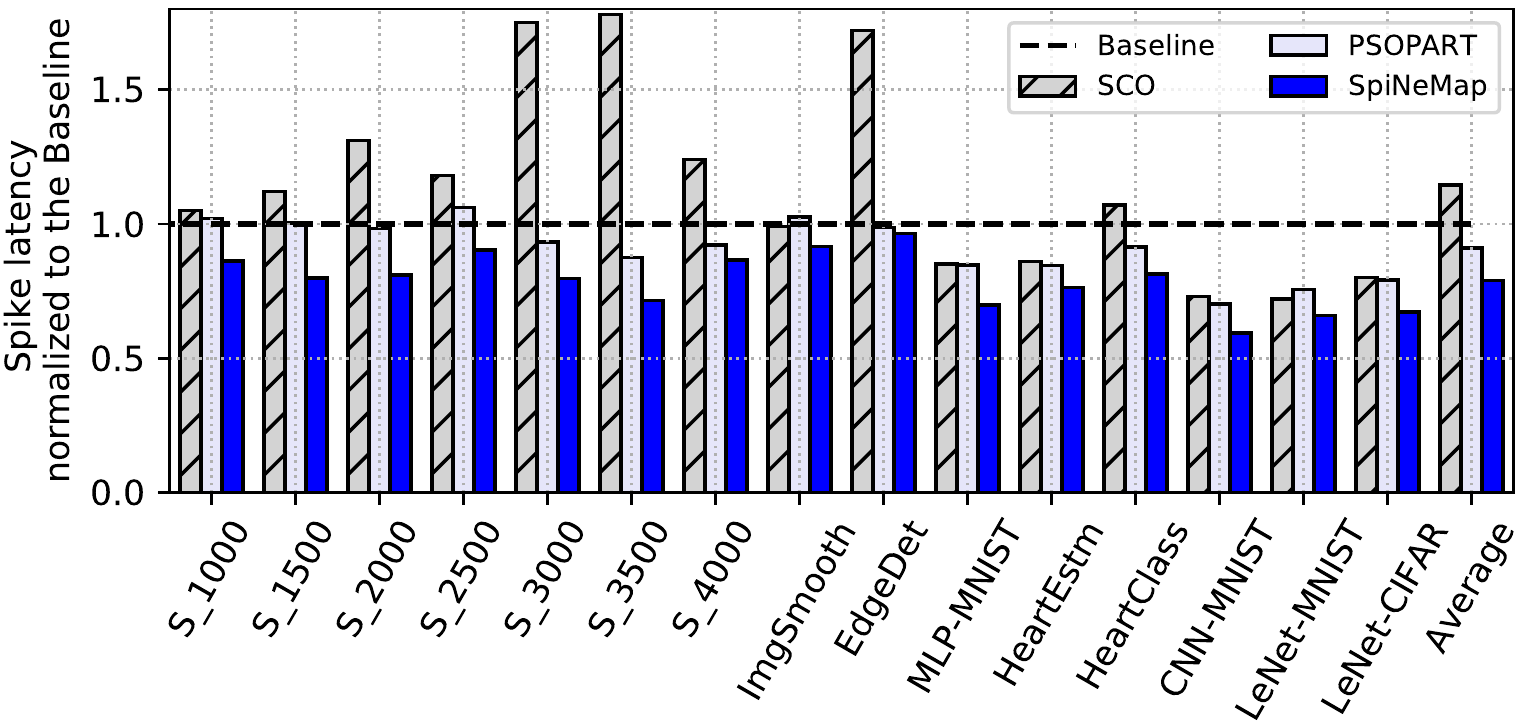}}
	\caption{Spike latency on the shared interconnect normalized to the Baseline.}
	\label{fig:latency}
\end{figure}

In Figure \ref{fig:latency}, we report the spike latency of the global synapses on the shared interconnect of each of our applications for each of our evaluated systems normalized to the Baseline. We make the following three observations.

\mr{
\textit{First}, the average spike latency of SCO is \ineq{14\%} higher than the Baseline. This increase is because SCO balances the crossbar utilization in the hardware and in doing so it can place certain synapses with large number of spikes on the shared interconnect, increasing the congestion and therefore the latency.
}
\textit{Second}, \ours{} has \ineq{9\%} lower average spike latency than the Baseline. This improvement is because \ours{} reduces the total number of spikes on the shared interconnect, which reduces spike congestion, improving the latency. 
\textit{Third}, \tech{} has the lowest average spike latency among all our evaluated systems (\LatencyImprovement{} lower average spike latency than the Baseline, \ineq{27\%} lower than SCO, and \ineq{13\%} lower than \ours{}).
\mr{These improvements are due to \tech{}'s optimization policies: 1) \techcluster{}, which reduces the number of spikes on the shared interconnect, reducing congestion and latency and 2) \techcluster{}, which minimizes latency by minimizing the average number of hop counts that spike traverses before reaching their destination.}

\subsection{Average ISI distortion of spikes on the shared interconnect}
\label{sec:isi}
\begin{figure}[h!]
	\centering
	\centerline{\includegraphics[width=0.99\columnwidth]{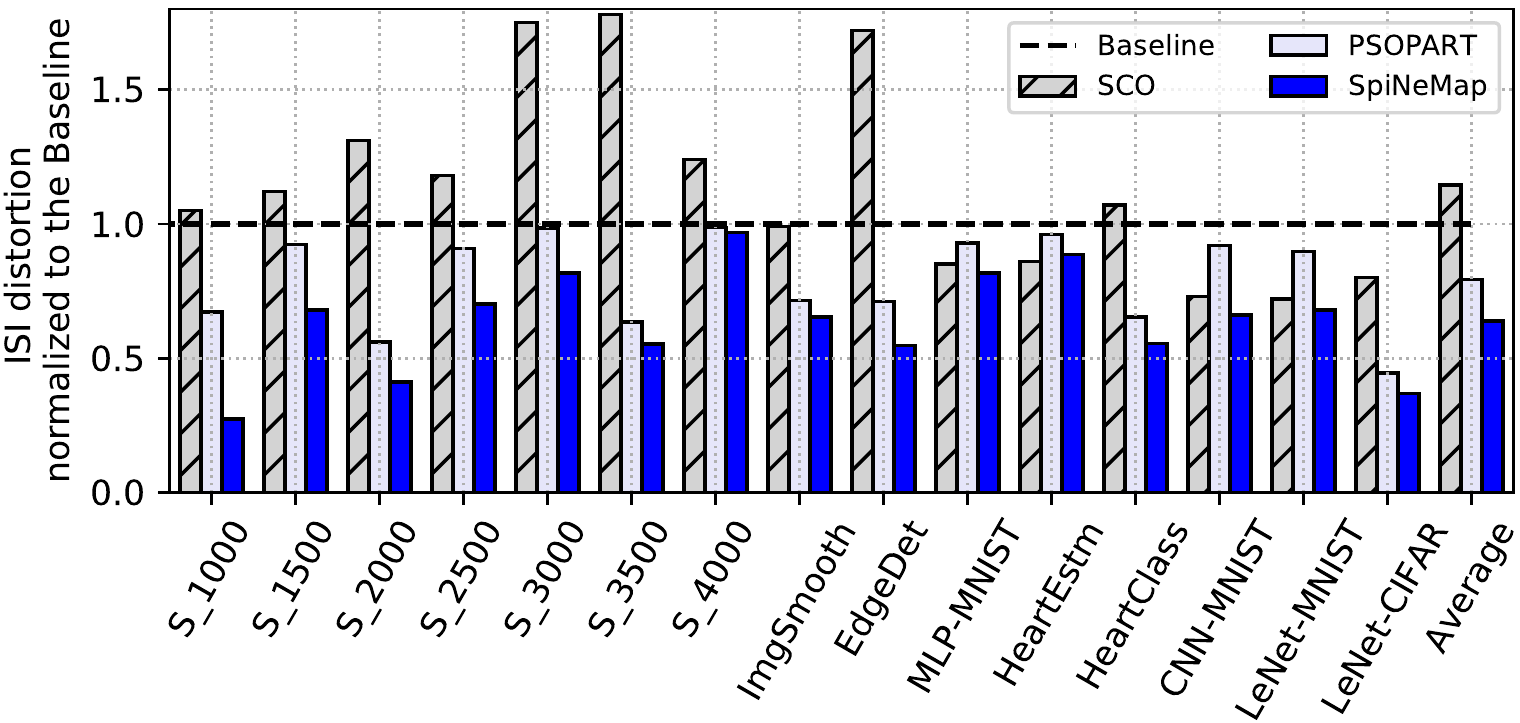}}
	\caption{Average ISI distortion normalized to the Baseline.}
	\label{fig:isi}
\end{figure}

In Figure \ref{fig:isi}, we compare the average inter-spike interval (ISI) distortion on the shared interconnect of each of our applications for each of our evaluated systems normalized to the Baseline. We make the following three observations.

\mr{\textit{First}, the ISI distortion of SCO is \ineq{12\%} higher than the Baseline. This increase is due to the increase in total spikes on the shared interconnect, which increases spike congestion and ISI distortion.
}
\textit{Second}, \ours{} has \ineq{21\%} lower average ISI distortion than the Baseline. This reduction is due to the reduction of the number of spikes on the shared interconnect. 
\textit{Third}, \tech{} has the lowest ISI distortion of all our evaluated systems (\isiImprovement{} lower average ISI distortion than the Baseline, \ineq{39\%} lower than SCO, and \ineq{23\%} lower than \ours{}). \mr{The improvement with respect to \ours{} is because of our new \techplacer{} step (see Figure \ref{fig:overview}), which further reduces the ISI distortion while reducing the spike latency.}

\subsection{Application accuracy}
\label{sec:app_accuracy}

\begin{figure}[h!]
	\centering
	\centerline{\includegraphics[width=0.99\columnwidth]{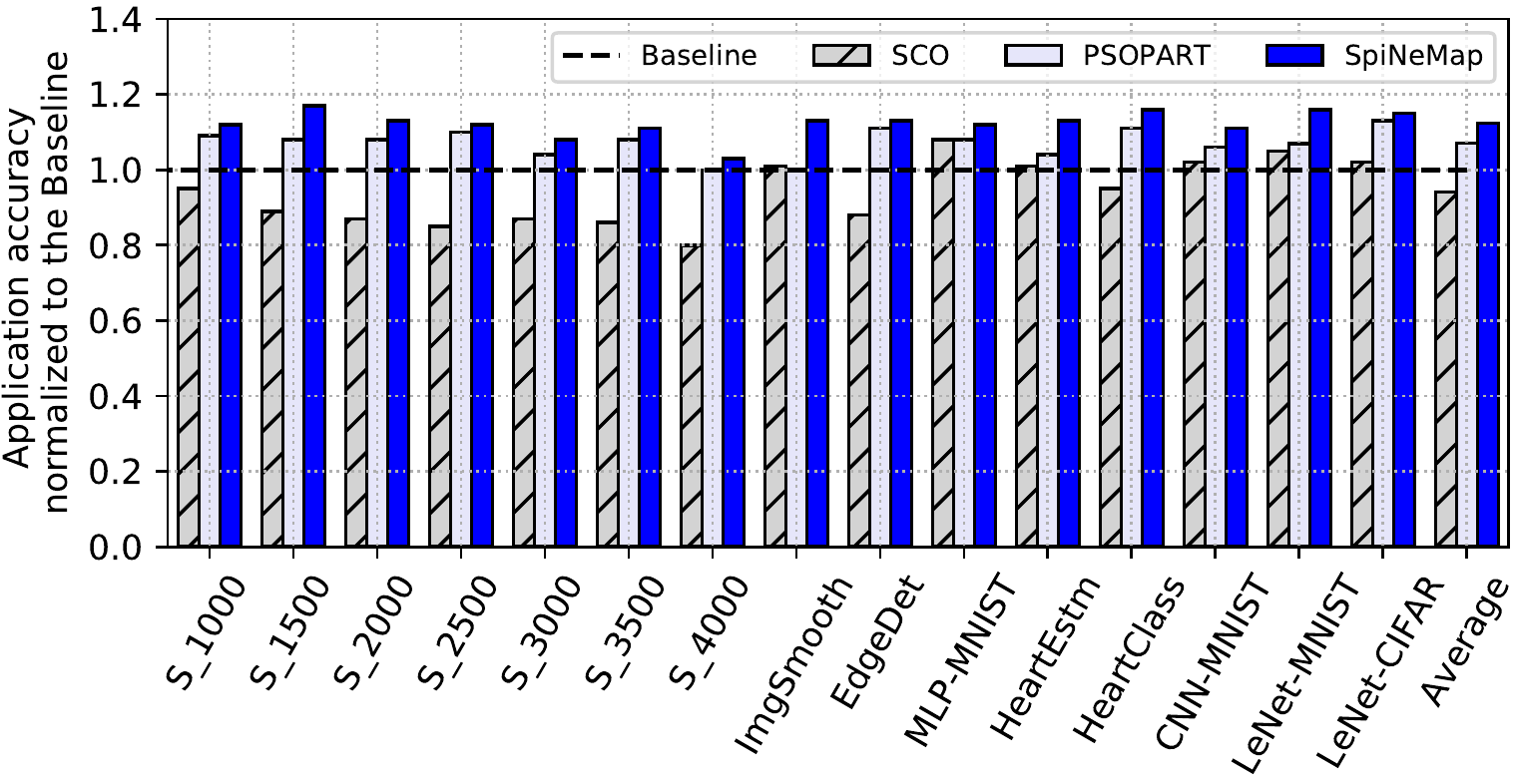}}
	\caption{Application accuracy normalized to the Baseline.}
	\label{fig:accuracy}
\end{figure}

In Figure \ref{fig:accuracy}, we report the application accuracy of each of our applications for each of our evaluated systems normalized to the Baseline. 
\mr{
We observe that the accuracy results directly correlate with the ISI distortion results we presented in Section \ref{sec:isi}. Specifically, the accuracy using SCO is lower than the Baseline by an average \ineq{6\%} due to the \ineq{12\%} increase in ISI distortion. \ours{} increases the accuracy by \ineq{7\%} due to the \ineq{17\%} reduction of ISI distortion. Finally, \tech{} achieves the highest accuracy among all our evaluated systems (\AccuracyImprovement{} higher average accuracy than the Baseline, \ineq{20\%} higher than SCO, and \ineq{5\%} higher than \ours{}).
}

\subsection{Evaluation of \techcluster{} in terms of spike count}
\label{sec:cluster_eval}
\begin{figure}[h!]
	\centering
	\centerline{\includegraphics[width=0.99\columnwidth]{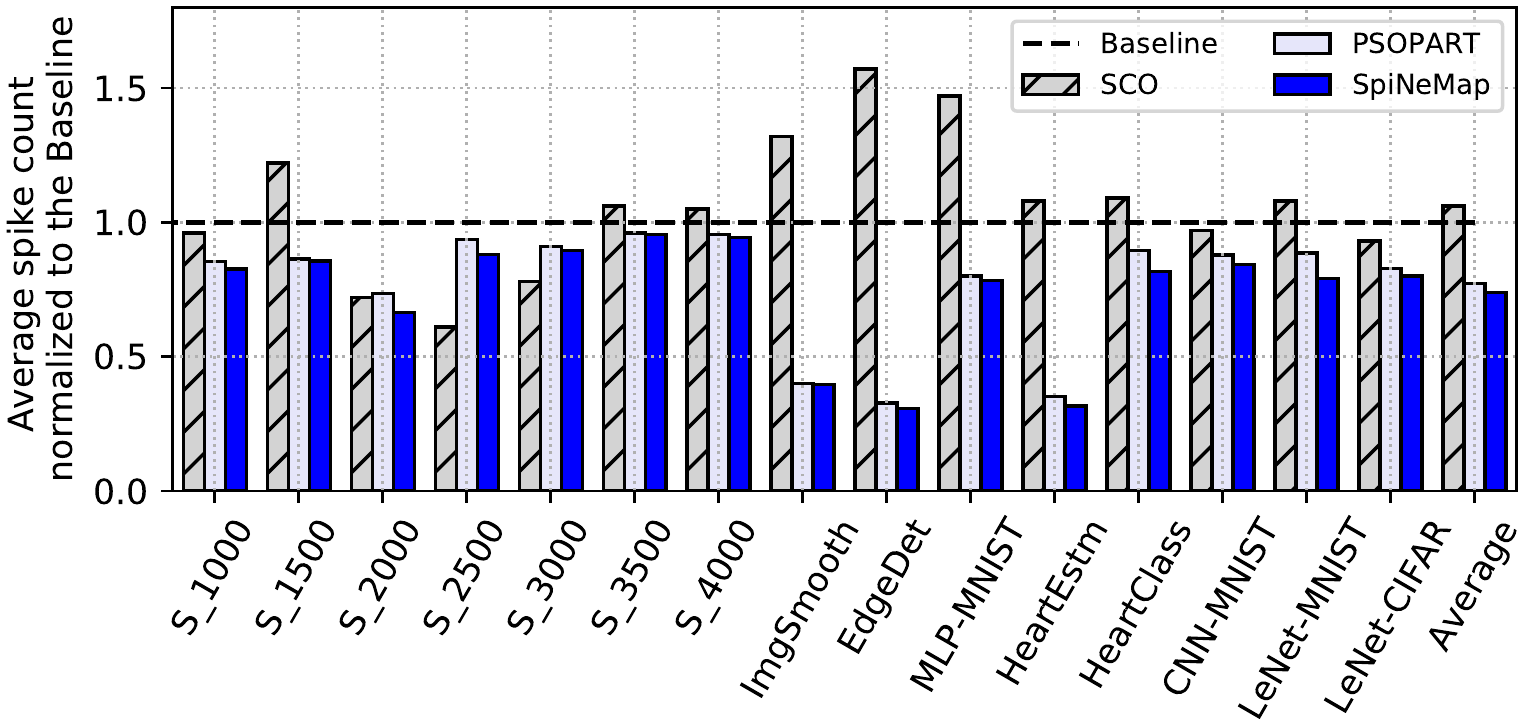}}
	\caption{Number of spike on the shared interconnect normalized to the Baseline.}
	\label{fig:cluster}
\end{figure}

In Figure \ref{fig:cluster}, we compare the total number of spikes communicated on the shared interconnect of each of our applications for each of our evaluated systems normalized to the Baseline. We make the following three observations.

\mr{
\textit{First}, SCO has an average \ineq{6\%} higher number of spikes on the shared interconnect compared to the Baseline. These extra spikes increases the energy consumption on the shared interconnect, which we presented in Section \ref{sec:energy_consumption}. \textit{Second}, \ours{} has \ineq{23\%} lower number of spikes due to the PSO approach, which explicitly minimizes the total number of spikes on the shared interconnect. \textit{Third}, \tech{} generates the lowest number of spikes on the shared interconnect (\ineq{26\%} lower than the Baseline, \ineq{24\%} lower than SCO, and \ineq{9\%} lower than \ours{}) The improvement over \ours{} is due to the greedy approach of Algorithm \ref{alg:two_part_full}, which outperforms the PSO, especially for the large application use-cases.
}

\subsection{Evaluation of \techcluster{} in terms of optimization time}
\label{sec:scalability}
\begin{figure}[h!]
	\centering
	\centerline{\includegraphics[width=0.99\columnwidth]{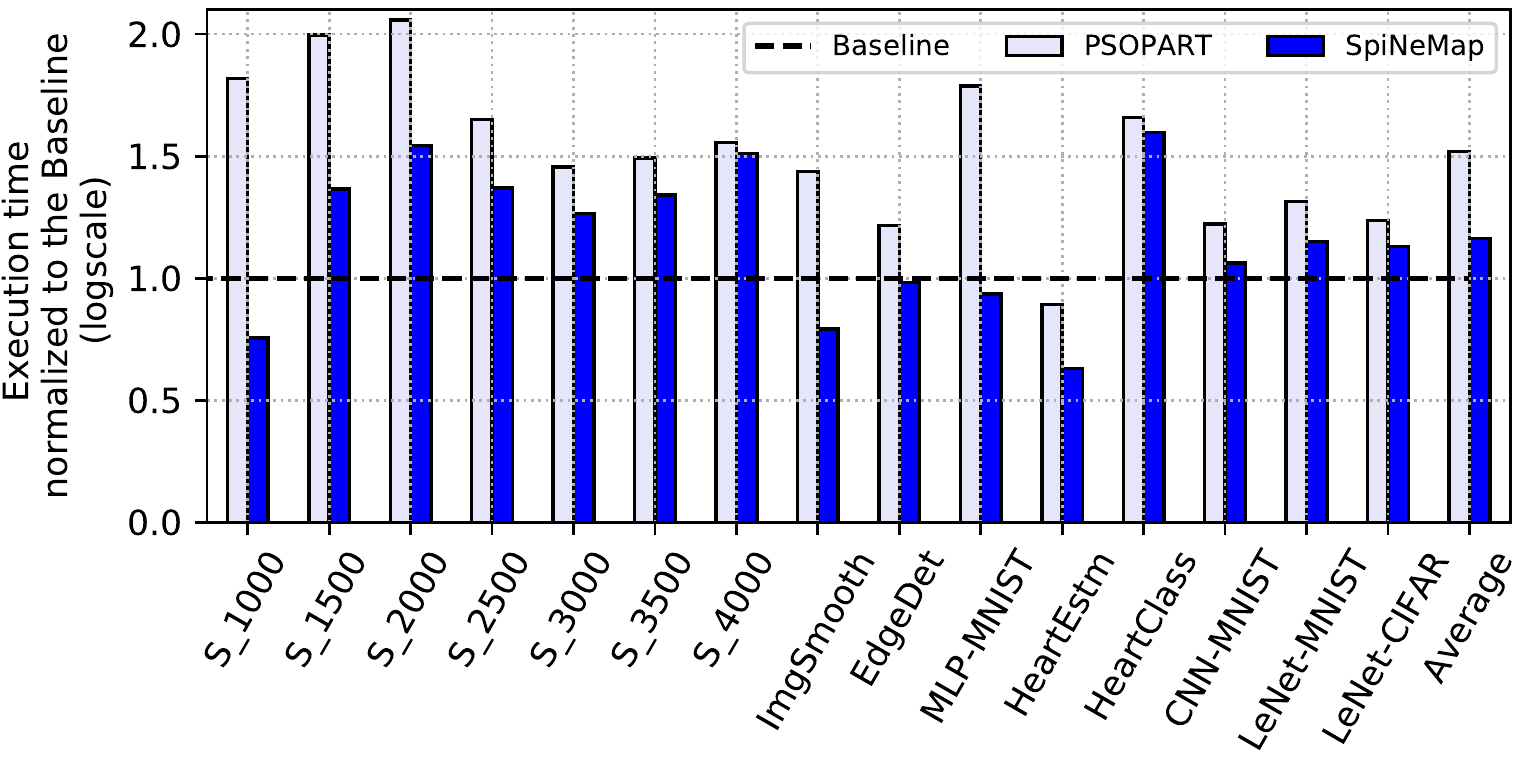}}
	\caption{Execution time normalized to the Baseline.}
	\label{fig:time}
\end{figure}

In Figure \ref{fig:time}, we compare the execution time of 
our new clustering algorithm (Algorithm \ref{alg:two_part_full}) against the PSO-based clustering
approach of \ours{}
normalized to the Baseline. 

\mr{
We observe that our \techcluster{} has an average 3x lower execution time than our previously-proposed PSO-based \ours{}. Additionally, we have shown in Section \ref{sec:cluster_eval} that Algorithm \ref{alg:two_part_full} generates an average 9\% lower number of spikes than the PSO-based solution, improving energy consumption, spike latency, and application accuracy. We conclude that our new clustering algorithm is scalable and generates better results than our previously-proposed PSO-based approach.
}

\subsection{\mr{Interconnect design explorations}}
\begin{figure}[h!]
	\centering
	\centerline{\includegraphics[width=0.99\columnwidth]{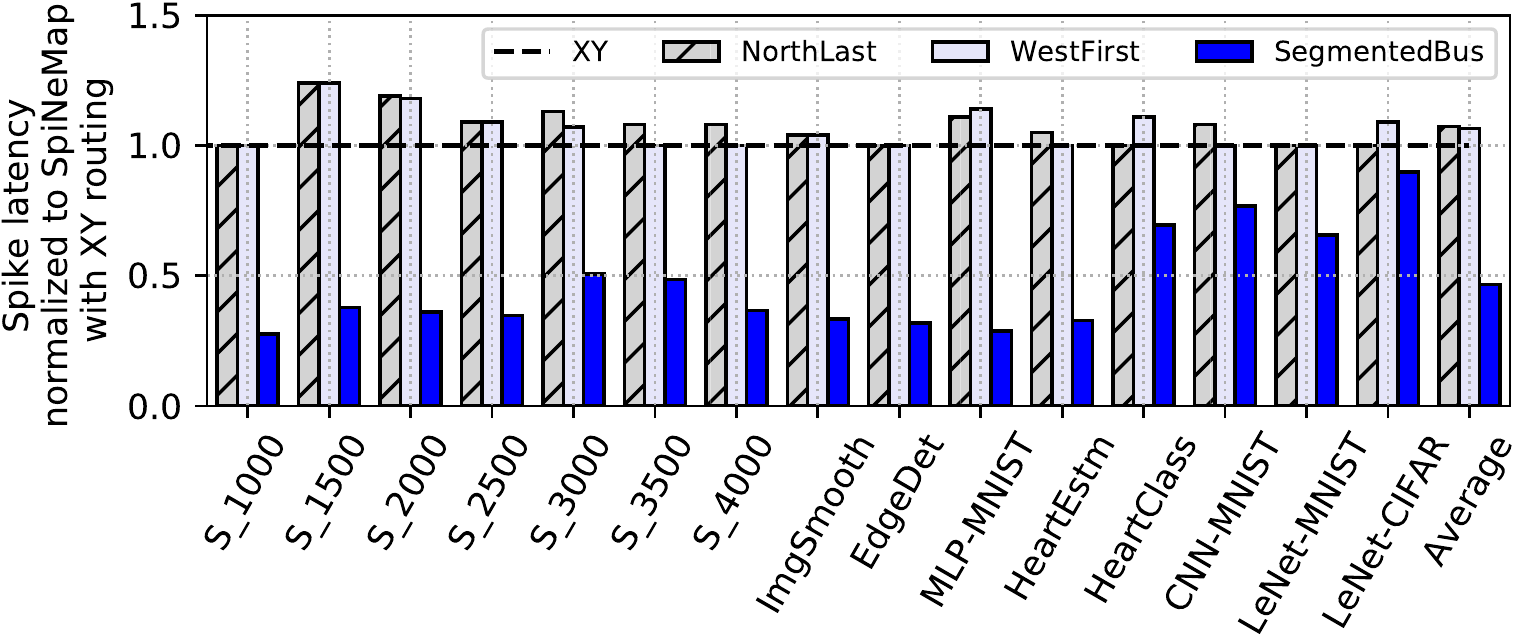}}
	\caption{Exploration of interconnects and routing algorithms using \tech{}}
	\label{fig:explorations}
\end{figure}

\mr{
In Figure \ref{fig:explorations}, we illustrate how our design methodology can be used for explorations on interconnect for neuromorphic hardware. In this figure, we compare XY routing, which is used in DynapSE against NorthLast and WestFirst routing algorithms. Finally, we evaluate our previously-proposed segmented bus \cite{Balaji2019ExplorationComputing} as an alternative to the multi-stage NoC used in the DynapSE neuromorphic platform. We evaluate these alternatives for all our evaluated workloads. We make the following two observations.
}

\mr{
\textit{First}, the NorthLast and WestFirst routing algorithms have an average \ineq{7\%} and \ineq{4\%} higher latency than the default XY routing algorithms, meaning that the XY routing algorithm is the most suitable one for the applications. \textit{Second}, the segmented bus interconnect has the lowest spike latency among all our evaluated routing algorithms (average 54\% lower for all these three routing algorithms). Lower spike latency leads to lower energy consumption and higher application performance.
}

\mr{
Our \tech{} design methodology allows simulating NoCs, segmented bus, and other interconnect topologies, facilitating future research on scalable interconnect for neuromorphic computing. Our continuing work is to extend the \tech{} with architecture of TrueNorth, Loihi, and other neuromorphic hardware platforms.
}
\section{Related Works}
\label{sec:related_works}
This is the first work that jointly addresses the partitioning and placement of SNNs on crossbar-based neuromorphic hardware, minimizing the energy consumption, spike latency, and ISI distortion, and improving application accuracy.

\subsection{SNN-based machine learning}
\mr{
Machine learning techniques such as neural networks \cite{LeCun2015DeepLearning} have proved to be immensely successful in many domains such as computer vision \cite{Szegedy2016RethinkingVision} and natural language processing \cite{Graves2013SpeechNetworks}.
The machine learning database MLPerf \cite{mlperf} provides a comprehensive collection of these applications. We demonstrate the performance of \tech{} using applications from MLPerf benchmark suite.
}
\mr{
Compared to analog and rate models, machine learning techniques implemented with spike model \cite{Brette2015PhilosophyBrain} and brain-inspired learning algorithms \cite{dan2004spike}, e.g., spiking neural networks \cite{maass1997networks},
have ultra-low power footprint when executed on neuromorphic hardware such as DYNAP-SE \cite{Moradi_etal18}, TrueNorth \cite{debole2019truenorth}, and Loihi \cite{davies2018loihi}. 
This makes spike-based computation model attractive for implementing machine learning applications on 
these devices.
}
Verstraeten et al. propose reservoir computing with SNNs for speech recognition \cite{verstraeten2006reservoir}. Grzyb et al. use spiking liquid state machine for facial recognition \cite{grzyb2009facial}.  Diehl et al. propose hand-written digit recognition using SNNs \cite{diehl2015unsupervised}. We have previously proposed a liquid state machine approach for heart-rate estimation from ECG signals \cite{das2018unsupervised}. \mr{We demonstrate the performance of \tech{} using some of these applications.}

\mr{
Recent works have demonstrated techniques to convert operations of analog computation model to spike model. One example is the N2D2 tool \cite{n2d2}. Using this tool we have previously demonstrated the SNN implementation of convolutional neural networks (CNN)-based heart-beat classification \cite{balaji2018power}.
}

\subsection{Neuromorphic hardware}
Recently, several research initiatives are undertaken to develop crossbar-based neuromorphic hardware using the emerging non-volatile memory technologies. Ramasubramanian et al. propose to use Spin-transfer torque magnetic RAM (STT MRAM) to build neuromorphic crossbars \cite{ramasubramanian2014spindle}. Burr et al. propose to use phase-change memories (PCM) to design neuromorphic crossbars \cite{burr2017neuromorphic}. Mallik et al. propose to use oxide-based resistive RAM (OxRAM) as alternative \cite{mallik2017design}. While all these orthogonal works focus on the design of a crossbar, we focus on the architecture of a neuromorphic chip integrating multiple such crossbars. To this end, Khan et al. propose a mapping strategy for SNNs on the SpiNNaker platform \cite{khan2008spinnaker}. Ji et al. propose NEUTRAMS for mapping neural networks on crossbar-based neuromorphic hardware \cite{ji2016neutrams}. In Section \ref{sec:results} we compare \tech{} against NEUTRAMS (i.e., the Baseline) and found that \tech{} is significantly better in terms of energy, latency, and application accuracy.
\subsection{SNN and neuromorphic simulators}
\tech{} is a technique that maps trained SNNs on the neuromorphic hardware. To this end, there are several choices for application-level SNN simulators that can generate trained SNNs. 
PyNN \cite{davison2009pynn} is a high level, simulator-independent interface used for building neuronal
models by providing high level abstractions allowing the access of low-level details
like neuron and synapse models of the computing back-end. There are also other simulators such as Brian \cite{goodman2009brian}, GeNN \cite{yavuz2016genn}, and NEST \cite{gewaltig2007nest}. We use CARLsim \cite{Chou2018CARLsim4} due to its detailed STDP and homeostasis models, parameter tuning, and multi-GPU support to accelerate the simulation. Nevertheless, \tech{} can be combined with any other SNN simulators.


\subsection{Related concepts in similar domains}
Graph partitioning problem has been extensively used for multiprocessor systems, where  an application task graph is partitioned to map tasks on the processing cores. The survey paper \cite{singh2013mapping} provides an overview of different mapping techniques and optimization objectives that have been proposed for multiprocessor systems.
These mapping techniques cannot be directly used for clustering because of the new metric ISI distortion that is specific to SNN. We chose the clustering technique in \techcluster{} because it is scalable and generates a good starting solution for the \techplacer{}.

\section{Conclusion and Future Outlook}\label{sec:conslusion}
\mr{
This paper introduces \tech{}, a design methodology to map SNN-based applications to crossbar-based neuromorphic hardware. \tech{} completes the mapping in two steps. In Step 1 (\techcluster{}), we use a heuristic-based clustering algorithm to partition SNNs into local and global synapses, with local synapses mapped within crossbars, and global synapses to the shared interconnect. Our objective is to minimize the number of spikes on the shared interconnect, which reduces spike congestion, leading to a reduction of the ISI distortion. In Step 2 (\techplacer{}), we use an instance of the particle swarm optimization (PSO)  to place clusters on physical crossbars in the hardware, optimizing energy consumption and spike latency on the shared interconnect. 
}

\mr{
Our optimization strategies in the two steps also improves application accuracy.
We evaluate \tech{} using synthetic and realistic SNN applications.
\tech{} reduces energy consumption on the shared interconnect by \EnergyImprovement{} and spike latency by \LatencyImprovement{}, compared to the state-of-the-art techniques. This reduces ISI distortion by \isiImprovement{}, which improves application accuracy by \AccuracyImprovement{} over state-of-the-art approaches.
}

\mr{
We {believe} that \tech{} is an end-to-end design methodology to map SNN applications on neuromorphic hardware. 
Our \tech{} framework is open-sourced and can be downloaded from the url \texttt{https://github.com/drexel-DISCO/SpiNeMap}.
}

\subsection{\mr{Future Outlook}}
\mr{
In this section, we describe how our design methodology \tech{} can be used to advance neuromorphic computing.
}

\subsubsection{\underline{\mr{Mapping new machine learning approaches to hardware}}} 
\mr{
Supervised machine learning approaches are usually limited when remembering and dealing with rare events.
Advanced machine learning approaches are therefore investigated. Many of these new proposals are based on spiking events. Examples include the liquid state machine \cite{maass2002real}, zero-shot learning \cite{Socher2013Zero-shotTransfer}, one-shot learning \cite{Fei-Fei2006One-shotCategories}, lifelong learning \cite{Silver2013LifelongAlgorithms}, transfer learning \cite{Pan2009ALearning}, and deep reinforcement learning \cite{Mnih2015Human-levelLearning} among others. All these new approaches can be mapped to hardware using \tech{}, by first simulating the application behavior in CARLsim, and then using the spike trace to partition and place the clusters on to hardware. In fact, in this work we demonstrate the mapping of one such emerging machine learning approach viz the liquid state machine implemented in the \textit{HeartEstm} application.
}

\mr{
From the computational neuroscience models front, we have demonstrated our design methodology \tech{} using the \textit{spike-based model}. Machine learning algorithms designed with the \textit{analog model} such as CNN or MLP can also be used in our design methodology by first converting the analog model to a spike-based model before presenting the application to \tech{}. In this work, we demonstrate this using three analog CNN-based applications. We converted these applications to spike-based model using the N2D2 tool \cite{n2d2}.
}

\mr{
For the \textit{rate model}, information is encoded as average firing rate of neurons in the SNN.
ISI distortion due to congestion on the interconnect does not always lead to performance loss as long as the average number of spikes received within a given time interval remains the same. 
A relevant metric for the rate model to capture the effect of spike congestion on the shared interconnect is the \textit{spike disorder}. We provide a proper intuition behind spike disorder as follows: We consider that a source neuron generates three spikes at time t = 0ns, 5ns, 25ns and 50ns. The spike rate of the source neuron are 200MHz and 50MHz, respectively. These three spikes need to be communicated to a destination neuron. 
We consider a scenario where spike 0 and 2 are received earlier at the destination neurons at time t = 5ns and 30ns, and spike 1 is re-routed due to congestion, reaching the destination neuron at t = 35ns. The spike rate received at the destination is therefore 40MHz and 200MHz, respectively. This spike disorder can lead to performance loss. 
We can formalize the definition of spike disorder as follows. Let \ineq{F^i = \{F_1^i,\cdots,F_{n_i}^i\}} be the expected spike arrival rate at neuron \ineq{i} (from CARLsim) and \ineq{\hat{F}^i = \{\hat{F}_1^i,\cdots,\hat{F}_{n_i}^i\}} be the actual spike rate considering hardware latencies. The spike disorder is computed as 
\begin{equation}
    \label{eq:spike_disorder}
    \footnotesize \text{spike disorder} = \sum_{j=1}^{n_i} [(F_j^i - \hat{F}_j^i)^2] / n_i
\end{equation}
}

\mr{
Our \techcluster{} can be trivially extended with minimum effort to compute and minimize spike disorder.
} 

\subsubsection{\underline{\mr{Using \tech{} for other neuromorphic platforms}}} 
\mr{
Our design methodology uses CARLsim to extract neural activity on every synapse of SNNs. CARLsim's support for built-in biologically realistic neuron, synapse, and computation models, designing new machine learning approaches and online learning algorithms,  
and continuous integration and testing, make it an easy to use and powerful simulator of biologically-plausible neural network models. The present release allows for the simulation using multiple GPUs and multiple CPU cores concurrently in a heterogeneous computing cluster. Benchmarking results demonstrate simulation of 8.6 million neurons and 0.48 billion synapses using 4 GPUs and up to 60x speedup for multi-GPU implementations over a single-threaded CPU implementation, making CARLsim 4 well- suited for large-scale SNN models in the presence of real-time constraints. Additionally, the present release adds new features, such as leaky-integrate-and-fire (LIF), 9-parameter Izhikevich, multi-compartment neuron models, and fourth order Runge Kutta integration.
}

\mr{
{\tech{} is a \textit{general-purpose design methodology} for mapping SNN-based applications to neuromorphic hardware. We have seamlessly integrated \tech{} with both open-sourced SNN simulators such as Brian \cite{goodman2009brian} and proprietary simulators such as XNet \cite{bichler2013design}.} As the input for \tech{} is the precise time of neural activity on every synapse, \tech{} can be extended with minimum effort to consider any SNN simulator that allows extracting spike timing information.
}

\mr{
Our \techplacer{} uses the Noxim \cite{catania2015noxim} simulator for cycle-accurate simulation of neuromorphic interconnect. To this end, we have previously evaluated many other simulators such as BookSim2 \cite{jiang2010booksim} and NIRGAM \cite{jain2007nirgam} for neuromorphic computing. Noxim allows significant advantage in terms of trace-driven simulations, extensions to other interconnect types, etc. See our prior work \cite{huynh2016exploration} for discussion of these alternatives. Our design-methodology \tech{} can be trivially extended to consider other interconnect simulators as long as they support 1) cycle-accurate simulation, and 2) trace-driven simulation. The former requirement is necessary to precisely compute the spike latency, which impacts performance (such as accuracy) of spike-based computation model. The second requirement is necessary to simulate application-level spike behavior in hardware considering delays on the interconnect.
}

\mr{
Finally, our \tech{} is demonstrated to work with the DynapSE neuromorphic hardware \cite{Moradi_etal18}. Our continuing work is to support other neuromorphic architectures including TrueNorth \cite{DeBole2019TrueNorth:Years} and Loihi \cite{davies2018loihi}. We have open-sourced our framework to foster future research in neuromorphic computing.
}


%




\ifCLASSOPTIONcaptionsoff
\fi



\bibliographystyle{IEEEtran}
\bibliography{snnhw,mendeley}
%
%
%

%

\vspace{-25pt}

\begin{IEEEbiographynophoto}{Adarsha Balaji}
	Adarsha Balaji received a Bachelor’s degree from ﻿Visvesvaraya Technological University, India, in 2012 and a Master's degree from Drexel University, Philadelphia, PA, in 2017. He is currently pursuing a Ph.D. degree from the Department of Electrical and Computer Engineering, Drexel University, Philadelphia, PA. His current research interests include design of neuromorphic computing systems, particularly data-flow and power optimization of spiking neural networks (SNN) hardware. 
\end{IEEEbiographynophoto}
\begin{IEEEbiographynophoto}{Anup Das}
	Dr. Anup Das is an Assistant Professor at Drexel University. He received a Ph.D. in Embedded Systems from National University of Singapore in 2014. Prior to his Ph.D., he was a research engineer for more than 7 years at ST Microelectronics (India and Grenoble) and LSI Corporation (India). Following his Ph.D., he was a post-doctoral fellow at the University of Southampton and a researcher at IMEC. His research focuses on neuromorphic computing and architectural exploration. He is a senior member of the IEEE.
\end{IEEEbiographynophoto}

\begin{IEEEbiographynophoto}{Yuefeng Wu}
	Yuefeng was enrolled in a joint master program of KTH, Royal Institute of Technology, Stockholm, Sweden and Technology University of Eindhoven after receiving his bachelor degree from Tianjin University. He worked at IMEC NL for his master thesis and researched on the communication mechanisms of neuromorphic computing. He designed and implemented the simulator for communication based on Noxim. He joined ING Groep N.V. as a management trainee in the track of IT after graduation and currently works as an information architect.
\end{IEEEbiographynophoto}

\begin{IEEEbiographynophoto}{Khanh Huynh}
	Biography not available.
\end{IEEEbiographynophoto}

\begin{IEEEbiographynophoto}{Francesco G. Dell'Anna}
	Francesco G. Dell'Anna was born in Gallipoli, Italy, on 14 January
	1993. He received the BE degree in computer engineering and the ME
	degree in embedded systems from Polytechnic of Turin in 2014 and 2016
	respectively. In 2016 he attended the electrical engineering master
	program at KULeuven, working on a neuromorphic simulator in IMEC
	(Belgium). He is currently a researcher in the Institute of Applied
	Micro-Nano Systems Technology, Key Laboratory of Micro-Nano System
	Technology and Smart Transduction, Chongqing Technology and Business
	University, Chongqing, China, and a Ph.D. student in the department of
	Micro- and Nanotechnology systems at the university college of
	southeast Norway. In 2018 he then joined Omnivision Technology as a
	Digital Designer in Oslo. His research interests include image
	sensors, neural networks, piezoelectric energy harvesters and low
	power electronic designs.
\end{IEEEbiographynophoto}

\begin{IEEEbiographynophoto}{Giacomo Indiveri}
	Giacomo Indiveri is a Professor at the Faculty of Science of the
	University of Zurich, Switzerland, director of the Institute of
	Neuroinformatics (INI) of the University of Zurich and ETH Zurich,
	and head of the Neuromorphic Cognitive Systems group at INI.
	Indiveri was awarded an ERC starting grant in 2011, and an ERC
	consolidator grant in 2017. He is interested in the study of real
	and artificial neural processing systems, and is building hardware
	neuromorphic cognitive systems, using full custom analog and
	digital VLSI technology. The circuits he develops are designed to
	emulate the physics of computation of biological neural processing
	systems and are aimed at building  autonomous  agents that can
	learn and reason about the actions to take in response to the
	combinations of external stimuli, internal states, and behavioral
	objectives. These "neuromorphic cognitive agents" are used to
	validate brain inspired computational paradigms in real-world
	scenarios, and to develop a new generation of fault-tolerant
	event-based neuromorphic computing technologies.
\end{IEEEbiographynophoto}

\begin{IEEEbiographynophoto}{Jeffrey L. Krichmar}
	Jeffrey L. Krichmar received a B.S. in Computer Science in 1983 from the University of Massachusetts at Amherst, a M.S. in Computer Science from The George Washington University in 1991, and a Ph.D. in Computational Sciences and Informatics from George Mason University in 1997. He spent 15 years as a software engineer on projects ranging from the PATRIOT Missile System at the Raytheon Corporation to Air Traffic Control for the Federal Systems Division of IBM. From 1999 to 2007, he was a Senior Fellow in Theoretical Neurobiology at The Neurosciences Institute. He currently is a professor in the Department of Cognitive Sciences and the Department of Computer Science at the University of California, Irvine. Krichmar has nearly 20 years experience designing adaptive algorithms, creating neurobiologically plausible network simulations, and constructing brain-based robots whose behavior is guided by neurobiologically inspired models. He has over 100 publications and holds 7 patents. His research interests include neurorobotics, embodied cognition, biologically plausible models of learning and memory, neuromorphic applications and tools, and the effect of neural architecture on neural function. He is a Senior Member of IEEE and the Society for Neuroscience.
\end{IEEEbiographynophoto}


\begin{IEEEbiographynophoto}{Nikil D. Dutt}
	Nikil D. Dutt (F) received a Ph.D. in Computer Science from the University of Illinois at Urbana-Champaign in 1989, and is currently a Distinguished Professor of
Computer Science, Cognitive Sciences, and EECS at the University of California, Irvine. He is also a Distinguished Visiting Professor in the CSE department at IIT Bombay, India. Dutt’s research interests are in embedded systems, electronic design automation (EDA), computer systems architecture and software, healthcare IoT, and brain-inspired architectures and computing. He received over a dozen best paper awards and nominations  at premier EDA and embedded systems conferences and is coauthor of 7 books on topics covering hardware synthesis, memory and computer architecture specification and validation, and on-chip networks. Dutt has served as Editor-in-Chief of ACM TODAES and as Associate Editor for ACM TECS and IEEE TVLSI.  He has extensive service on the steering, organizing, and program committees of several premier EDA and Embedded System Design conferences and workshops, and also serves or has served on the advisory boards of ACM SIGBED, ACM SIGDA, ACM TECS, IEEE Embedded Systems Letters (ESL), and the ACM Publications Board.

	He is a Fellow of the ACM, Fellow of the IEEE, and recipient of the IFIP Silver Core Award.

\end{IEEEbiographynophoto}

\begin{IEEEbiographynophoto}{Siebren Schaafsma}
	Dr. Siebren Schaafsma is an R\&D manager in the IoT unit of Imec – The Netherlands (Imec-nl). This unit is part of the Holst Center in Eindhoven. He is responsible for two teams of Analog and Digital IC designers building new state of the art Radio IC’s and sub GHz Radar (BT-LE, Wifi 11.ah, subGHz, etc). He is also responsible for a team of embedded hardware and software engineers working in the domain of IoT and Artificial Intelligence. He received two masters (Nuclear physics in 1988 and computer science in 1989) at the Rijks Universiteit Groningen (RUG). His dissertation in the latter one addresses a neural networks implementation on a transputer cluster. He received his Ph.D. (Dr.) from the University of Nijmegen (KUN) in the Biophysics Department. His dissertation addresses the coding of optic flow in the visual cortex. He holds two patents on his research inventions from his period in research at Ericsson Telecommunications.
\end{IEEEbiographynophoto}

\begin{IEEEbiographynophoto}{Francky Catthoor}
	Dr. Francky Catthoor received  a Ph.D. in EE from the Katholieke Univ. Leuven,
	Belgium in 1987.  Between 1987 and 2000,  he  has headed  several research
	domains in the area of synthesis techniques and architectural methodologies.
	Since 2000 he is  strongly involved in other activities at IMEC including
	deep submicron technology aspects, IoT and biomedical platforms, and
	smart photovoltaic modules, all at IMEC Leuven,  Belgium.   Currently he is
	an IMEC fellow.
	He is also part-time full professor at the EE department of the KULeuven.
	
	He has been associate editor for several IEEE and ACM journals.
	He was elected IEEE fellow in 2005.
\end{IEEEbiographynophoto}

%
%
%




\end{document}